\documentclass[useAMS,usegraphicx,usenatbib]{mn2e}

\usepackage{txfonts,amssymb,hyperref} 

\newcommand{\Msunyr}{\hbox{${\rm M}_\odot\,{\rm yr}^{-1}$}}

\newcommand{\zem}{\hbox{$z_{\rm em}$}}

\newcommand{\EBV}{\hbox{$E(B\!-\!V)$}}
\newcommand{\apropto}{\,\rlap{\raise 0.4ex\hbox{$\propto$}}{\lower
    0.5ex\hbox{$\sim$}}\,}

\newcommand{\bspsmall}{\vspace{0.5cm}\small\noindent This paper has been typeset
from a \TeX/\LaTeX\ file prepared by the author.\normalsize}

\defcitealias{Pettini_Pagel_2004a}{PP04}
\defcitealias{Wild_etal_2006a}{WHP06}
\defcitealias{Wild_etal_2007a}{WHP07}

\def\CaII{Ca\,{\sc ii}}

\def\Ha{\rm{H}\alpha}
\def\Hb{\rm{H}\beta}
\def\MgI{Mg\,{\sc i}}
\def\MgII{Mg\,{\sc ii}}
\def\NaI{Na\,{\sc i}}
\def\FeI{Fe\,{\sc i}}

\def\SiII{Si\,{\sc ii}}
\def\ZnII{Zn\,{\sc ii}}
\def\HI{H\,{\sc i}}
\def\HII{H\,{\sc ii}}
\def\NII{[N\,{\sc ii}]}
\def\OII{[O\,{\sc ii}]}
\def\OIII{[O\,{\sc iii}]}

\def\Lya{\rm{Ly}\alpha}
\def\zabs{z_{\rm abs}}
\def\zem{z_{\rm em}}

\def\IRAF{{\sc iraf}}
\def\UVESpopler{{\sc uves popler}}

\title[\CaII\ absorption-selected galaxies at $z<0.5$]{The host galaxies of
  strong \CaII\ QSO absorption systems at $\bmath{z<0.5}$}

\author[B.~J.~Zych et al.]
       {Berkeley~J.~Zych,$^1$\thanks{E-mail:~\href{mailto:bjz@ast.cam.ac.uk}{bjz@ast.cam.ac.uk} (BJZ)}
	 Michael~T.~Murphy,$^1$ Max~Pettini,$^1$ Paul~C.~Hewett,$^1$	 
	 \newauthor Emma~V.~Ryan-Weber$^1$ and Sara~L.~Ellison$^2$\\
	 $^1$Institute of Astronomy, University of Cambridge, Madingley Road, Cambridge CB3 0HA\\
	 $^2$Dept. Physics \& Astronomy, University of Victoria, 3800 Finnerty Road, Victoria, BC, V8P 1A1, Canada
       }

\begin{document}

\date{Accepted ---. Received ---; in original form ---}
\pagerange{\pageref{firstpage}--\pageref{lastpage}}
\pubyear{2007}

\maketitle

\label{firstpage}

\begin{abstract}
  We present new imaging and spectroscopic observations of the fields
  of five QSOs with very strong intervening \CaII\ absorption systems
  at redshifts $z_{\rm abs} < 0.5$ selected from the Sloan Digital Sky
  Survey. Recent studies of these very rare absorbers indicate that
  they may be related to damped Lyman alpha systems (DLAs). In all
  five cases we identify a galaxy at the redshift of the \CaII\ system
  with impact parameters up to $\sim\!24\,\rm{kpc}$. In four out of
  five cases the galaxies are luminous ($L \simeq L^{\ast}$),
  metal-rich ($Z \simeq Z_{\odot}$), massive (velocity dispersion
  $\sigma \simeq 100\,$km~s$^{-1}$) spirals.  Their star formation
  rates, deduced from H$\alpha$ emission, are high, in the range ${\rm
  SFR} = 0.3 - 30\,\Msunyr$.  In our analysis, we paid particular
  attention to correcting the observed emission line fluxes for
  stellar absorption and dust extinction. We show that these effects
  are important for a correct SFR estimate; their neglect in previous
  low-$z$ studies of DLA-selected galaxies has probably led to an
  underestimate of the star formation activity in at least some DLA
  hosts. We discuss possible links between \CaII-selected galaxies and
  DLAs and outline future observations which will help clarify the
  relationship between these different classes of QSO absorbers.
\end{abstract}


\section{Introduction}

Following star formation throughout the Universe's history is key to
understanding how galaxies have evolved.  A wide variety of methods
have been used to probe galaxies and their star formation activity
over the cosmic ages.  At high redshifts, most of the information at
our disposal comes on the one hand from direct observations of
star-forming galaxies over most of the electromagnetic spectrum --
from X-ray to radio wavelengths -- and, on the other, from indirect,
but more precise, studies of the interstellar absorption lines
produced by intervening galaxies in the spectra of background quasars
(QSOs).  These two techniques should be complementary, and yet it has
proved difficult until recently to bring their respective results
together into one comprehensive picture.  The aim of this paper is to
bridge this gap by presenting \emph{direct} imaging and spectroscopic
observations of galaxies originally selected via their absorption
characteristics.
 
The class of QSO absorption-line systems (QALs) which has been studied
most extensively in this respect are the damped Lyman-$\alpha$ systems
(DLAs), defined by neutral hydrogen column densities $N_{{\rm
H}\,\textrm{\sc i}} \geq 2.0\times10^{20}\,\rm{atoms}\,\rm{cm}^{-2}$.
At the upper of the column density distribution of QALs, DLAs should
account for most of the neutral gas in the Universe at $z\sim 3$ and
have thus long been regarded as the likely reservoirs of gas available
for star formation \citep*{Wolfe_etal_2005a}.  However, this potential
for star formation is apparently yet to be realized by most galaxies
identified as DLAs.  Attempts to measure the star-formation rate of
DLA hosts -- whether through direct imaging
(\citealt{Wolfe_Chen_2006a}; \citealt*{Hopkins_etal_2005a}) and
spectroscopic detection \citep[and references
therein]{Kulkarni_etal_2006a}, or through indirect arguments based on
the inferred heating rate \citep*{Wolfe_etal_2003a} -- have generally
returned relatively modest values, lower than expected from a
straightforward application of the Schmidt law to their surface mass
densities of \HI.  Similarly, the metal content of most DLAs is less
than 1/10 solar at most redshifts \citep{Akerman_etal_2005a,
Kulkarni_etal_2005a}, lower than that of the disk of the Milky Way
over most of its past history \citep{Pettini_2006a}.

These characteristics are in stark contrast with those of galaxies
detected directly through their starlight at similar redshifts in
surveys such as those of
\citet{Steidel_etal_2003a,Steidel_etal_2004a}.  Such `Lyman break', or
more generally UV-bright, galaxies have star-formation rates ${\rm
SFR} \simeq 10-100\,\Msunyr$
\citep[e.g.][]{Reddy_etal_2005a,Reddy_etal_2006a} and have already
reached near-solar metallicities at $z = 2-3$
\citep{Pettini_etal_2001a,Erb_etal_2006c}.

If we are to reconcile these apparently contradictory results, it is
essential to image the DLA absorbers and measure the metallicities of
their star-forming regions, for comparison with the values measured in
the cold gas which evidently dominates the absorption cross-section.
Clearly, such efforts should focus first on DLAs at $z \lesssim 1$,
where both imaging and spectroscopy are easier to perform than at
higher redshifts.  Once the connection between galaxies and DLAs has
been clarified at $z \lesssim 1$, it may be easier to interpret the
data at higher redshifts. While there have been a few attempts in this
general direction (e.g. \citealt*{Ellison_etal_2003a};
\citealt{Bowen_etal_2005a}; \citealt*{Chen_etal_2005a};
\citealt{Zwaan_etal_2005a}), the body of relevant data is still very
thin, mainly because surveys of DLAs at $z < 1$ require ultraviolet
spectroscopy from space which is currently unavailable until the next
servicing mission installs the Cosmic Origin Spectrograph on the
\emph{Hubble Space telescope} (\emph{HST}).

Until then we have to rely on methods other than the direct detection
of a damped Lyman-$\alpha$ absorption line to identify DLAs at $z
\lesssim 1$.  We are greatly aided in such endeavours by the
unprecedented statistical power of the \textit{Sloan Digital Sky
Survey} (SDSS).  In a recent development, \citet{Wild_Hewett_2005a}
identified a sample of rare QALs characterized by strong
\CaII\,$\lambda\lambda$3934,3969 absorption.\footnote{These are the
familiar Ca\,{\sc ii} K \& H lines first identified in the solar
spectrum and subsequently studied extensively in the interstellar
medium of the Milky Way.}  \citet{Wild_Hewett_2005a} showed that
Ca\,{\sc ii} systems selected to have a rest-frame equivalent width of
the stronger member of the doublet $W_0^{3934} \geq 0.2\,\rm{\AA}$ are
likely to have high values of $N_{{\rm H}\,\textrm{\sc i}}$; they may
well be a subclass of DLAs \citep*[henceforth
\citetalias{Wild_etal_2006a}]{Wild_etal_2006a}. However, their
connection to the Ly$\alpha$-selected DLAs has yet to be fully
clarified. \citetalias{Wild_etal_2006a} also found that the $z\sim1$
strong \CaII\ absorbers have higher average dust extinction than most
DLAs [$\EBV\sim0.1\,\rm{mag}$] and average dust-to-metals ratio as
high or higher than the Milky Way. Their number density was also found
to be $\sim$20--30 per cent that of the DLA population at the same
redshift. From recent $K$--band imaging of strong \CaII\ absorbers,
also at $z\sim 1$, \cite{Hewett_Wild_2007b} find a mean impact
parameter of $24\,\rm{kpc}$ with a filling factor of only $\sim$10 per
cent for their sample. They also find that the luminosity dependence
of the \CaII\ absorber cross-section [$\sigma \propto
\left(L/L^\ast\right)^{0.7}$] is stronger than the dependence
established for strong \MgII\ absorbers.

From the extensive body of work carried out on \CaII\ absorption in
the Galactic interstellar medium (ISM) over the last 50 years, we do
know that, locally at least, the \CaII\ lines are seldom very strong
because: (i) with an ionization potential close to, but lower than,
that of \HI\ (11.9\,eV compared to 13.6\,eV), \CaII\ may not be the
major ionization stage of Ca in \HI\ regions where a significant
fraction of Ca may be doubly ionized; and (ii) more importantly, Ca is
among the most depleted elements in the gaseous phase of the ISM,
being readily incorporated into dust grains
\citep[e.g.][]{Savage_Sembach_1996a}. Thus, it is conceivable that
strong \CaII\ absorption may preferentially arise in environments
where some fraction of the grains has been destroyed -- and the
proportion of gaseous Ca consequently enhanced by a large factor -- by
supernova-driven shocks associated with star-formation activity
\citep{Routly_Spitzer_1952a}.

Partly to test this hypothesis, \citet*[henceforth
\citetalias{Wild_etal_2007a}]{Wild_etal_2007a} stacked the SDSS
spectra of QSOs with strong \CaII\ systems to search for \OII\
emission from the absorbing galaxies. They found that the population
as a whole exhibits only modest levels of in-situ star formation
activity, with an average rate $\left<\rm{SFR}\right> = 0.11 -
0.48\,\Msunyr$. In order to place this result in context, establish
the nature of the strong \CaII\ absorbers, and clarify their
relationship, if any, to the DLA population, we have begun a programme
of deep imaging and spectroscopy targeting initially the fields of
SDSS QSOs with strong \CaII\ absorption at $z_{\rm abs} \lesssim
0.5$. In this paper we report the first results of this survey for
five fields: in each case we have detected galaxies at the redshift of
the absorber and, unlike previous studies, we find them to be
luminous, actively star-forming, and rich in metals and dust. While
this may be partly due to the bias of our first observations in favour
of easily recognized, and therefore luminous, galaxies, these initial
results show that \CaII\ absorbers should play an important role in
improving our understanding of the general absorber population in the
future.

The paper is structured as follows. In Section \ref{sec:data} we
review our sample selection, describe the observations, which were
conducted with the FORS2 spectrograph on the Very Large Telescope
(VLT) of the European Southern Observatory, and discuss the subsequent
data reduction steps.  We pay particular attention to fitting the
underlying stellar absorption along the Balmer series and to dust
extinction corrections.  In Section \ref{sec:caii_gen_results} we
describe the general properties of our \CaII-selected galaxies.  In
Section \ref{sec:Z_SFR_caii} we deduce star formation rates and oxygen
abundances using well established emission line diagnostics and
compare them with analogous measures from the literature for the hosts
of Ly$\alpha$-selected DLAs.  Finally, Sections \ref{sec:discussion}
and \ref{sec:conclusions} present our discussion and conclusions,
respectively.

Throughout this paper we adopt the currently favoured values of the
cosmological parameters $H_0 = 70$\,km~s$^{-1}$~Mpc$^{-1}$,
$\Omega_{\rm M,0} = 0.3$ and $\Omega_{\Lambda,0} = 0.7$.  Wavelengths
are quoted in a vacuum, heliocentric frame of reference.

\section{Sample selection, observations, data reduction and analysis}\label{sec:data}
In addition to describing the observations, data reduction and
analysis, we pay particular note to the systematic effects involved in
our analysis. This includes fitting the underlying stellar absorption
along the Balmer series and dust reddening corrections.

\subsection{Selection}
Absorbers were identified by searching for features with rest frame
equivalent width, $W_0^{3934}\geq 0.2\,\rm{\AA}$ at $4\,\sigma$
significance in the sample of all 49409 SDSS DR3 QSOs with $\zem \geq
0.05$, classified by having `specClass' $=3$ or $4$ in the SDSS. It
was also required that the \CaII\,$\lambda3969$ line had a
significance $>1\,\sigma$.  Detections were required to lie outside
the $\Lya$ forest and if two detections were separated by less than
$500\,\rm{km}\,\rm{s}^{-1}$ in redshift space they were classified as
a single absorption system.  Absorber redshifts were measured using
Gaussian profile fits. For a candidate absorber to be confirmed as
real it also had to show \MgII\ absorption with $>6\,\sigma$
significance at $\zabs\geq0.37$ or, for $\zabs<0.37$, \NaI\ absorption
at $>1\,\sigma$ since the wavelengths of the \MgII\ lines then lie
outside the SDSS spectral range. The sample was restricted to
$\zabs<0.5$ to allow the $\Ha$ emission line to be observed in the
optical and to allow easier identification of host galaxies from SDSS
imaging.  Finally, the sample spectra were visually inspected, to
reject absorbers which did not look convincing. This left a sample of
40 absorbers. Despite these checks there is a possibility that up to
ten per cent of the absorbers in the sample could be spurious due to
the difficulty of identifying the relatively weak \CaII\ features,
particularly at low-$z$ when stellar photospheric \CaII\ H\&K lines
are often detected.  Photospheric \CaII\ H\&K detections can usually
be eliminated by their very strong \CaII\ equivalent widths combined
with presence of other strong photospheric absorption features such as
the \NaI, \FeI\ and \MgI\ lines. However, some \CaII\ absorption was
subsequently found to be of stellar, rather than interstellar, origin
from FORS2 observations detailed in Section \ref{sec:Obs}. These
systems were removed from the sample and are not discussed in this
paper. Where possible, comparison between $W_0^{3934}$ measured in
SDSS and, less contaminated, FORS2 QSO spectra offers another method
of ensuring the interstellar origin of the absorber (see Table
\ref{tab:sources}).

\begin{figure*}
\begin{minipage}[c]{1.0\textwidth}
\centering
\includegraphics[origin=c,scale=1.0,trim=0 0 0 0,angle=0,clip=true]{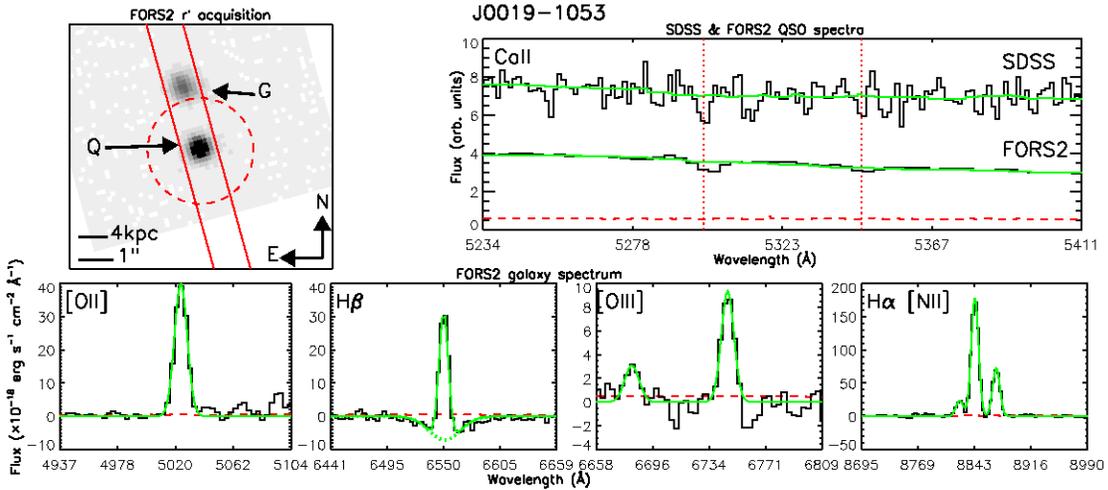}
\end{minipage}
\caption{Summary of the data available for J0019$-$1053. The top left
panel shows the stacked $r$--band FORS2 acquisition image. The solid
red lines shows the FORS2 slit positioning, whilst the dashed red line
shows the SDSS fibre placement. Image scale and orientation are
indicated in the bottom left and right hand corners, respectively. The
top right panel shows the \CaII\ line detection in the SDSS QSO
spectrum. We also plot the more significant FORS2 QSO \CaII\ line
detection (when available). For clarity, the FORS2 spectrum flux scale
is offset from that of the SDSS spectrum by four arbitrary flux
units. The solid black line shows the observed flux, whilst the solid
green and dashed red lines show the 41 pixel median filter continuum
and error array, respectively. The positions of the
\CaII$\,\lambda\lambda$3934,3969 lines are indicated by the dotted red
lines. The bottom panels show portions of the FORS2 galaxy spectrum
around the emission lines labeled in the top left corner of each
plot. The solid black line indicates the observed flux after galaxy
continuum subtraction. The solid green line shows the fitted Gaussian
profiles, whilst the dashed red line shows the error array in each
case. In the case of $\Hb$ the dotted green line shows the base of the
stellar absorption fit. Wavelength scales are in the observed frame.}
\label{fig:J001946}
\end{figure*}

\begin{figure*}
\begin{minipage}[c]{1.0\textwidth}
\centering
\includegraphics[origin=c,scale=0.50,trim=0 0 0 0,angle=0,clip=true]{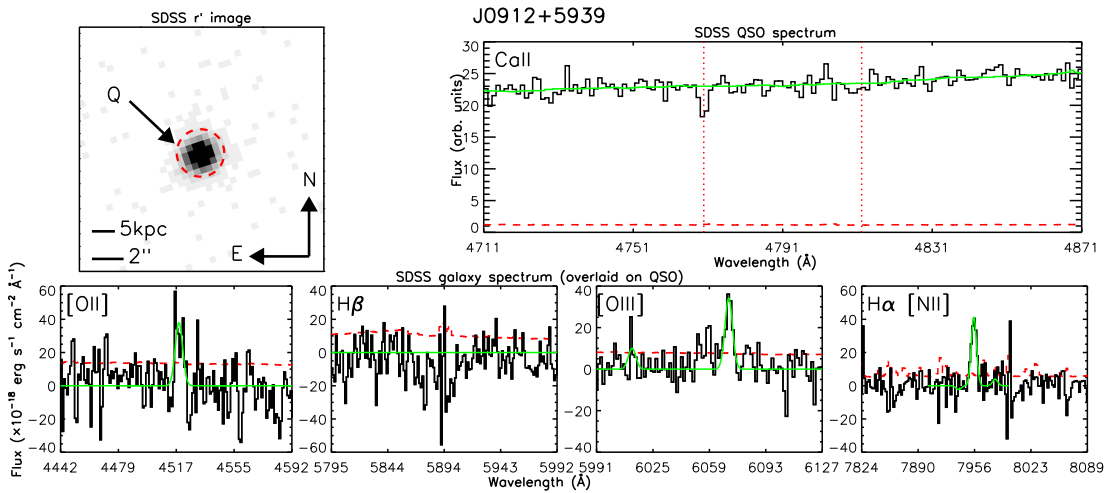}
\end{minipage}
\caption{Summary of the data available for J0912$+$5939. See
Fig.~\ref{fig:J001946} for description of each panel. Note that no
FORS2 galaxy spectra were taken so we present instead the SDSS
$r$--band image. The galaxy spectrum is observed through the same
fibre as the QSO. Note also that no absorption or emission fit is
possible to the $\Hb$ line; see Section \ref{sec:extinction}. The SDSS
MJD, fibre and plate numbers for this QSO are 51907, 459 and 0484,
respectively.}
\label{fig:J091204}
\end{figure*}

\begin{figure*}
\begin{minipage}[c]{1.0\textwidth}
\centering
\includegraphics[origin=c,scale=1.00,trim=0 0 0 0,angle=0,clip=true]{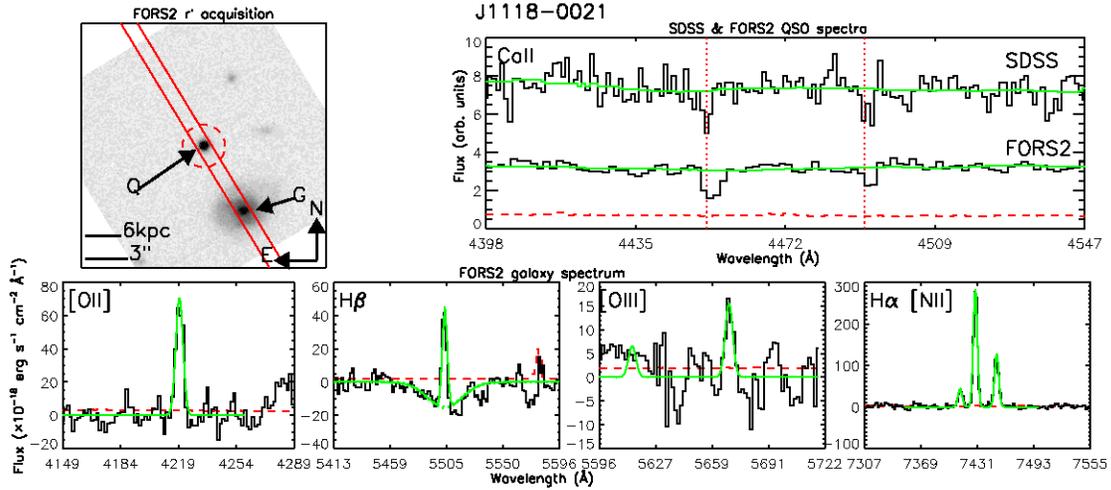}
\end{minipage}
\caption{Summary of the data available for J1118$-$0021. See
Fig.~\ref{fig:J001946} for description of each panel.}
\label{fig:J111850}
\end{figure*}

\begin{figure*}
\begin{minipage}[c]{1.0\textwidth}
\centering
\includegraphics[origin=c,scale=1.0,trim=0 0 0 0,angle=0,clip=true]{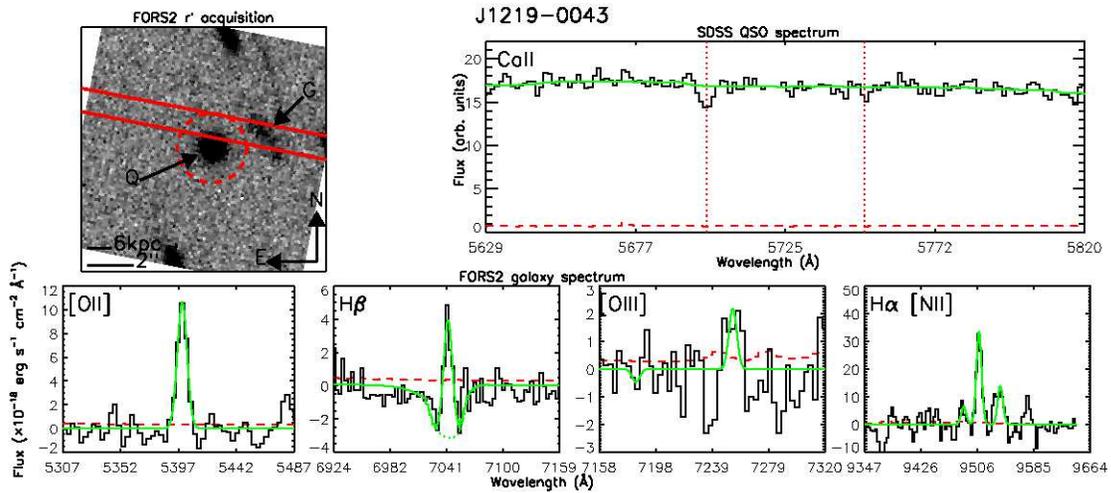}
\end{minipage}
\caption{Summary of the data available for J1219$-$0043. See
Fig.~\ref{fig:J001946} for description of each panel. Note that the
$\Hb$ absorption was fitted by eye in this case; see
Section \ref{sec:general_anal}.}
\label{fig:J121911}
\end{figure*}

\begin{figure*}
\begin{minipage}[c]{1.0\textwidth}
\centering \includegraphics[origin=c,scale=1.0,trim=0 0 0 0,angle=0,clip=true]{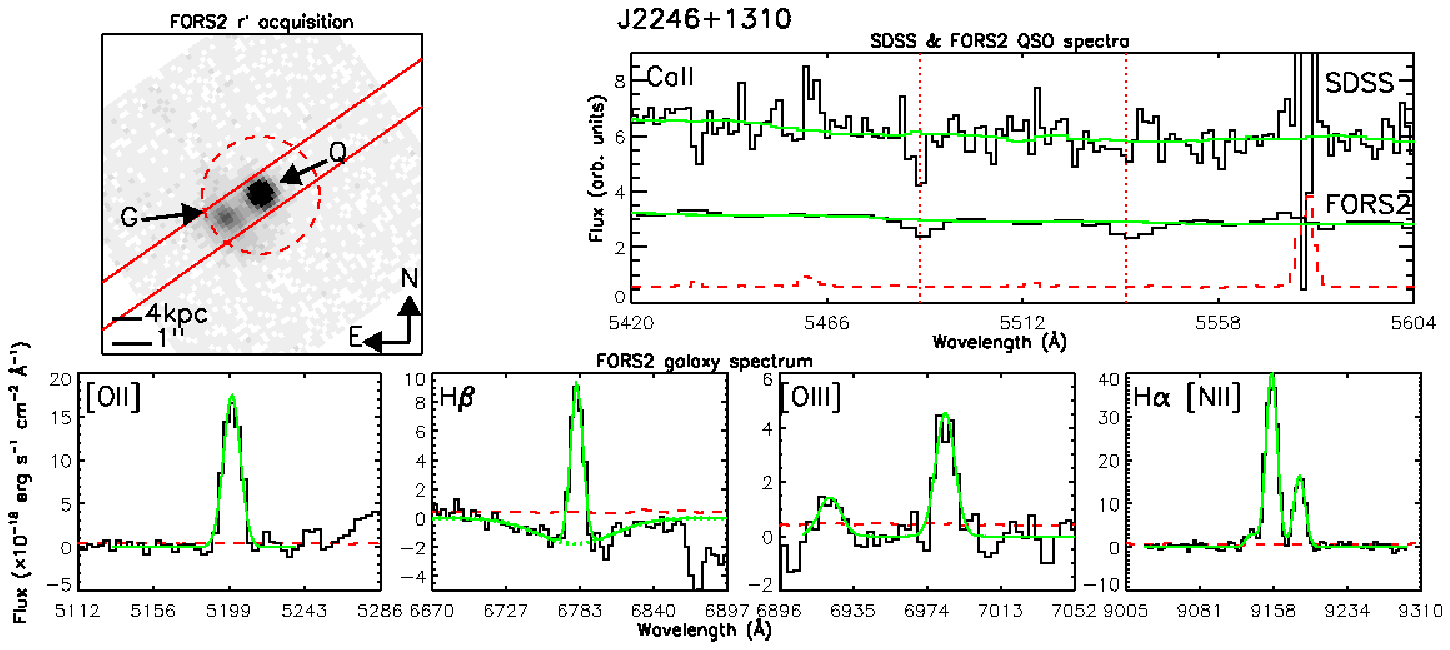}
\end{minipage}
\caption{Summary of the data available for J2246$+$1310. See
Fig.~\ref{fig:J001946} for description of each panel. For clarity, the
FORS2 spectrum flux scale is offset from that of the SDSS spectrum by
three arbitrary flux units.}
\label{fig:J224630}
\end{figure*}

The SDSS images were then examined by eye to identify likely host
galaxies in each field. Four sight lines, with particularly obvious
choices for host galaxies were chosen for follow up spectroscopy. Two
of the targets already had SDSS galaxy spectra, one observed through
the QSO fibre (J0912$+$5939), the other observed through a separate
fibre (J1118$-$0021), but we wished to improve upon the S/N of the
SDSS data for the latter. J0912$+$5939 is at a declination not
accessible from the VLT, but we analyse its SDSS galaxy spectrum later
in this paper.  Table~\ref{tab:sources}\ contains a list of the
targets examined. The SDSS names refer to the QSO J2000 coordinates,
rather than the galaxy.  The redshifts specified in this table refer
to the redshifts measured from the SDSS QSO CaII H\&K absorption
lines.  The final columns give the rest frame equivalent width of the
$\lambda$3934 line of each \CaII\ absorber as measured in their SDSS
and, where possible, FORS2 spectra, respectively.
Figs.~\ref{fig:J001946}--\ref{fig:J224630} show the collective data
for each galaxy, including the detected \CaII\ absorber from the SDSS
QSO spectrum and the FORS2 $r$--band acquisition images for each
galaxy, where available, otherwise we present the SDSS $r$--band
image. The orientation, fibre and slit positions are indicated on each
image.

\begin{table*}
\begin{minipage}[c]{0.8\textwidth}
\centering
\caption{Observed objects, redshifts and \CaII\ equivalent widths. The
  available galaxy spectral data sets are also noted
  (${^\dagger}$FORS2, $^{\ast}$SDSS)}
\label{tab:sources}
\begin{tabular}{lccccc}
  \hline
  Object & SDSS name & $z_{\rm abs}$ & $z_{\rm QSO}$ & $W_0^{3934} \left(\rm{SDSS}\right) /\,\rm{\AA}$ & $W_0^{3934} \left(\rm{FORS2}\right) /\,\rm{\AA}$\\[1.0ex]
  \hline
  J0019$-$1053$^{\dagger}$     & SDSSJ$001946.99\!-\!105313.3$ & 0.347 & 1.518 & $0.6 \pm 0.1$ & $0.47 \pm 0.03$\\
  J0912$+$5939$^{\ast}$        & SDSSJ$091204.90\!+\!593957.7$ & 0.212 & 0.773 & $0.52 \pm 0.09$ & --\\
  J1118$-$0021$^{\dagger\ast}$ & SDSSJ$111850.13\!-\!002100.8$ & 0.132 & 1.025 & $0.62 \pm 0.15$ & $0.79 \pm 0.03$\\
  J1219$-$0043$^{\dagger}$     & SDSSJ$121911.23\!-\!004345.5$ & 0.448 & 2.293 & $0.57 \pm 0.08$ & --\\
  J2246$+$1310$^{\dagger}$     & SDSSJ$224630.63\!+\!131048.5$ & 0.395 & 1.593 & $0.9 \pm 0.2$ & $0.59 \pm 0.02$\\
  \hline
\end{tabular}
\end{minipage}
\end{table*}


\subsection{Observations}\label{sec:Obs}
We obtained low resolution (R$\sim$600) spectra of four \CaII\
absorption-selected galaxies during six half nights of observations
using the FORS2 spectrograph at the VLT (see Table~\ref{tab:obs}).
The observations were split into two runs, one in 2006 April/May (run
A) and one in 2006 June (run B). Both runs had seeing
$\sim$0\farcs5-1\farcs5, but run A was significantly affected by cloud
such that only a few exposures could be reliably flux calibrated. This
was not a significant problem because other exposures were then scaled
to the absolute flux scale obtained with the observations taken in
clear conditions (see Section \ref{sec:reduction}).

The FORS2 spectrograph was used in long slit mode with a $1\arcsec$
slit and $2\times2$ CCD binning. The atmospheric dispersion corrector
(ADC) was utilized at all times and we observed all our galaxies at
airmass$<$1.5. Two combinations of grism and filter settings were
used, depending on the redshift of the galaxy being observed. The
first combination was 600B+12 (no filter, R$\sim$780) in the blue and
600RI+19 (filter GG435+81, R$\sim$1000) in the red. The second was
300V+20 (filter GG435+81, R$\sim$440) and 600z+23 (filter OG590+32,
R$\sim$1390). The settings were selected to ensure coverage of all the
emission lines of interest -- $\Ha$, \NII, $\Hb$, \OIII\ and \OII\ --
and to allow accurate relative flux calibration between red and blue
grism settings by virtue of their spectral overlap.

Exposures were typically $600\,\rm{s}$ in duration and multiple ($>$3)
exposures of each target were used to help eliminate cosmic rays. The
total time on each target was between 1 and $3\,\rm{hrs}$.  Standard
bias, flat field and flux calibration frames were also taken.

\subsection{Data reduction}\label{sec:reduction}\label{sec:absfluxcal}\label{sec:relfluxcal}
The FORS2 data were reduced using standard \IRAF\footnote{\IRAF\ is
distributed by the National Optical Astronomy Observatories, which is
operated by the Association of Universities for Research in Astronomy
Inc. (AURA) under cooperative agreement with the National Science
Foundation.} packages, except for the {\sc apextract} package which we
modified as detailed below. The 2D spectral frames were bias
subtracted, flat fielded and wavelength calibrated. Cosmic rays were
then identified via Laplacian edge detection using {\sc l.a. cosmic}
\citep{Van_Dokkum_2001a}. The modified version of {\sc apextract} was
then used to extract the 1D spectra.

We found that the standard \IRAF\ optimal extraction routine often
created inappropriate weighting profiles as our galaxy sources are
extended, rather than point-like. Therefore we added the option of
pure variance weighting of the data during extraction (i.e. using a
uniform spatial profile for weighting). Furthermore, we wished to
improve upon the cosmic ray identification and removal in
\IRAF. Therefore, we added the option of inputting a predetermined
cosmic ray mask into the extraction procedure. During extraction
\IRAF\ calculates the variance-weighted mean flux in the spatial
direction for each spectral pixel. We simply allowed pixels which were
pre-identified as cosmic rays to be masked during this process.

Finally, the 1D spectra were flux calibrated and corrected for
telluric absorption using observations of standard stars. Individual
spectra were vacuum and heliocentric corrected and combined using the
software package, \UVESpopler\footnote{Available at
\href{http://www.ast.cam.ac.uk/~mim/UVES\_popler.html}{http://www.ast.cam.ac.uk/$\sim$mim/UVES\_popler.html}.}. During
this process, all data were rebinned into the same linear wavelength
scale. Each exposure is also scaled up to the flux level of the
highest signal-to-noise ratio (S/N) exposure using a single scale
factor in each case. This ensures that, even for exposures which were
affected by cloud, we should achieve optimal flux calibration.

Between sets of exposures taken in photometric conditions we found
differences of less than five per cent in absolute flux
calibration. Good quality spectrophotometric standards and use of the
ADC ensure that relative flux calibration error is minimal -- the red
and blue flux distribution of individual exposures in each target show
that relative errors are up to four per cent. The SDSS data utilized
has accurate spectrophotometric calibration
\citep{Tremonti_etal_2004a}, thus flux calibration errors are not the
dominant source of error in our sample (see Section
\ref{sec:systematics}).

Emission lines were corrected for Galactic extinction using the Milky
Way (MW) dust map created by \citet*{Schlegel_etal_1998} and the
reddening curve given by \citet{Cardelli_etal_1989a}.

\subsection{Aperture corrections}\label{sec:geocor}
Geometric corrections are required for each galaxy to account for the
fact that the slit used for spectroscopy did not include all the
star-forming light from the galaxy. To calculate the corrections for
our sample we assume that our $r$--band acquisition imaging traces the
$\Ha$ and \OII\ line flux. The resulting corrections, derived by
taking the ratio of total galaxy $r$--band flux to that in the slit
are factors between 1 and 2.2. Individual corrections are given in
Table~\ref{tab:geocor} with the resulting corrected SFRs in
Section \ref{sec:CaII_SFRs}. We utilize these geometric corrections where
appropriate. In particular we use {\em uncorrected} SFRs when
comparing our work to that conducted on DLAs in the literature because
the latter measurements do not include any such corrections.

\subsection{General spectral analysis and removal of underlying stellar absorption}\label{sec:general_anal}

The data were analysed using a combination of the Starlink package,
{\sc dipso}, and {\sc ngaussfit} from the \IRAF\ {\sc stsdas}
package. The resolution of the data allowed the $\Ha$ and \NII\
emission lines to be resolved and accurate fitting of the stellar
absorption around $\Hb$. Continua were fitted interactively, using a
spline interpolation, and subtracted from the total flux. The emission
lines were fitted with Gaussian profiles to calculate total line flux.

There can be a large amount of uncertainty in the fitting of stellar
absorption along the Balmer series
\citep[e.g.][]{Diaz_1988a,Gonzalez_etal_1999a}; this in turn can cause
a significant uncertainty in line measurements and hence in the
estimated dust reddening from the Balmer decrement. We therefore
compared two different methods for fitting the $\Hb$ line, one fitting
the emission and absorption simultaneously using Gaussian profiles,
the other using a Gaussian profile in emission whilst the absorption
was spline fitted by eye.  The $W_0$ of the absorption is smaller when
fitted by eye, resulting in a ten per cent systematic difference in
the integrated line fluxes. The Gaussian absorption fits better
reflect the shape of the stellar absorption derived from stellar
synthesis models, built up from individual stars, than the by-eye
spline fits hence we adopted those where possible. J1219$-$0043 is the
exception to this, which was spline fitted by eye because the results
from {\sc ngaussfit} were very sensitive to the initial conditions for
the fit.

In most analyses in the literature the effects of stellar absorption
on $\Ha$ are neglected because the emission equivalent width is much
greater than that of absorption. However, $\Ha$ absorption was
significant for some of our objects. Note that $\Ha$ stellar
absorption can not be corrected directly because the \NII\ lines mask
the absorption at the resolution of our data. We therefore used the
following prescription to correct the $\Ha$ line for stellar
absorption. For normal star-forming spiral galaxies the following
relationship between $W_0$ of stellar absorption at $\Hb$ and $\Ha$
applies (J.~Moustakas, private communication),
\begin{equation}
\label{eqn:HaHbrel}
  W_0^{{\rm abs}}\left({\rm H}\alpha\right) = 0.23 + 0.59\times W_0^{{\rm abs}}\left({\rm H}\beta\right)\,.
\end{equation}
This relationship is based on a linear bisector fit to the measured
$\Ha$ and $\Hb$ stellar absorption from stellar synthesis modeling of
the galaxies in the \citet*{Moustakas_etal_2006a} and Nearby Field
Galaxy Survey \citep[NFGS,][]{Jansen_etal_2000a} samples. The
1$\sigma$ dispersion in the relationship is $0.11\,{\rm dex}$
($0.09\,{\rm dex}$, rejecting the ten 3$\sigma$ outliers from the
total sample of 407).

We then use the reasonable assumption that the velocity structures (of
both emission and absorption) at $\Hb$ and $\Ha$ are the same. This
allows us to correct the $\Ha$ line flux for stellar absorption via
\begin{equation}
\label{eqn:Hacor}
  F_{\rm abscor}\left({\rm H}\alpha\right) = \frac{W_0^{\rm em}\left({\rm H}\alpha\right) + f W_0^{\rm abs}\left({\rm H}\alpha\right)}{W_0^{\rm em}\left({\rm H}\alpha\right)}F_{\rm obs}\left({\rm H}\alpha\right)\,,
\end{equation}
where $W_0^{\rm abs}\left({\rm H}\alpha\right)$ is the equivalent
width derived in equation (\ref{eqn:HaHbrel}) and $W_{\rm r}^{\rm
em}\left({\rm H}\alpha\right)$ is the measured rest-frame equivalent
width of the $\Ha$ emission line. $f$ is a correction factor which
varies between spectra. It represents the fraction of the total
stellar absorption measured when calculating the equivalent width
across the emission profile, rather than across both the emission
profile and broader absorption profile. This is necessary because one
can not measure $W_0^{{\rm em}}\left({\rm H}\alpha\right)$ across the
same width in wavelength space as $W_0^{{\rm abs}}\left({\rm
H}\beta\right)$, due to the presence of the \NII\ lines (refer to
$\Ha$ and $\Hb$ spectra in
Figs.~\ref{fig:J001946}--\ref{fig:J224630}). However, it is
straightforward to measure $f$ directly from the $\Hb$ line.
Table~\ref{tab:Hbstellarabs} gives the measured $\Hb$ and $\Ha$
stellar absorption equivalent widths and the measured correction
factor, $f$. Noise in their data precluded this analysis on
J0912$+$5939 and J1219$-$0043, thus we do not correct their $\Ha$ line
fluxes and quote results based on them as lower limits. It is clear
that the corrections are not inappropriate given that our different
metallicity determinations all agree within $0.3\,\rm{dex}$ (see
Section \ref{sec:caii_Z}). Extracted spectra of the emission line
regions with fits used to calculate integrated line fluxes can be
found in Figs.~\ref{fig:J001946}--\ref{fig:J224630}\ and
Table~\ref{tab:lineflux}\ lists measured line fluxes.

\begin{table}
\caption{A summary of the observations taken at the ESO VLT using the
FORS2 long slit spectrograph.}
\label{tab:obs}
\begin{tabular}{lcccc}
\hline
Object & Grism & Filter & R & Exposure time / s\\[0.5ex]
\hline
J0019$-$1053 & 300V+20  & GG435+81 & 440  & 5200\\
             & 600z+23  & OG590+32 & 1390 & 2200\\   
J1118$-$0021 & 600B+12  & none     & 780  & 3600\\ 
             & 600RI+19 & GG435+81 & 1000 & 1200\\
J1219$-$0043 & 300V+20  & GG435+81 & 440  & 6540\\
             & 600z+23  & OG590+32 & 1390 & 3600\\   
J2246$+$1310 & 300V+20  & GG435+81 & 440  & 6400\\
             & 600z+23  & OG590+32 & 1390 & 2800\\   
\hline
\end{tabular}
\end{table}

\begin{table}
\caption{The stellar absorption as measured at $\Hb$ and derived for
$\Ha$ for our sample of galaxies. The correction factor, $f$, is
required in equation (\ref{eqn:Hacor}) to correct for stellar
absorption at $\Ha$. Also tabulated is the measured emission line
equivalent width at $\Ha$.}
\label{tab:Hbstellarabs}
\begin{tabular}{lccccc}
\hline
Object & $W_0^{{\rm abs}}\left(\Hb\right)\,/\,\rm{\AA}$ & $W_0^{{\rm abs}}\left(\Ha\right)\,/\,\rm{\AA}$ & $f$ & $W_0^{{\rm em}}\left(\Ha\right)\,/\,\rm{\AA}$\\[0.5ex]
\hline
J0019$-$1053	& $7 \pm 1$     & $4.2 \pm 0.7$ & 0.62 & $58.77 \pm 0.01$\\   
J1118$-$0021	& $4.6 \pm 0.8$ & $2.9 \pm 0.5$ & 0.31 & $14.825 \pm 0.002$\\ 
J2246$+$1310	& $5 \pm 1$     & $3.2 \pm 0.8$ & 0.31 & $28.81 \pm 0.02$\\   
\hline
\end{tabular}
\end{table}

\begin{table*}
  \begin{minipage}{1.0\textwidth}
    \centering
    \caption{Measured galaxy integrated line fluxes, Galactic dust
    extinctions and integrated galaxy dust extinctions for each
    target. We note that the systematic errors in \EBV$_{\rm gal}$ are
    approximately an order of magnitude larger than the quoted random
    errors and are dominated by the estimate of the stellar absorption
    at $\Hb$ and $\Ha$; see text. Upper limits are quoted at 3$\sigma$
    significance.}
    \label{tab:lineflux}
    \label{tab:extinctions}
    \begin{tabular}{lcccccccccc}
      \hline
      &\multicolumn{7}{c}{Integrated Line Flux / $\times 10^{-18}\,{\rm erg}\,{\rm s}^{-1}\,{\rm cm}^{-2}$}\\
      \raisebox{1.0ex}{Object} & $z_{\rm abs}$ & $\Ha$ & \NII\,$\lambda6585$ & \NII\,$\lambda6549$ & $\Hb$ & \OIII\,$\lambda5008$ & \OIII\,$\lambda4960$ & \OII\,$\lambda3727$ & \raisebox{1.0ex}{\EBV$_{\rm Gal}$} & \raisebox{1.0ex}{\EBV$_{\rm gal}$}\\[0.5ex]
      \hline
      J0019$-$1053 & 0.347 & $2230 \pm 40$ & $870 \pm 30$   & $280 \pm 30$  & $420 \pm 50$ & $100 \pm 7$  & $37 \pm 9$  & $460 \pm 20$ & $0.033 \pm 0.001$ & $0.55 \pm 0.04$\\
      J0912$+$5939 & 0.212 & $320 \pm 40$  & $<$110         & $<$40         & --          & $200 \pm 30$ & $60 \pm 20$ & $230 \pm 40$ & $0.042 \pm 0.001$ & --            \\
      J1118$-$0021 & 0.132 & $1730 \pm 20$ & $730 \pm 20$   & $250 \pm  20$ & $300 \pm 30$ & $80 \pm 20$  & $28$        & $390 \pm 30$ & $0.050 \pm 0.003$ & $0.61 \pm 0.04$\\
      J1219$-$0043 & 0.448 & $310 \pm 20$  & $130 \pm 20$   & $60 \pm 20$   & $83 \pm 8$   & $16 \pm 6$   & $<$5        & $101 \pm 8$  & $0.032 \pm 0.001$ & --            \\
      J2246$+$1310 & 0.395 & $610 \pm 10$  & $239 \pm 9$    & $43 \pm 9$    & $131 \pm 5$  & $58 \pm 6$   & $21 \pm 6$  & $213 \pm 7$  & $0.051 \pm 0.002$ & $0.42 \pm 0.02$\\
      \hline
    \end{tabular}
  \end{minipage}
\end{table*}

\begin{figure}
  \includegraphics[origin=c,width=\columnwidth,trim=0 0 0 0,angle=0,clip=true]{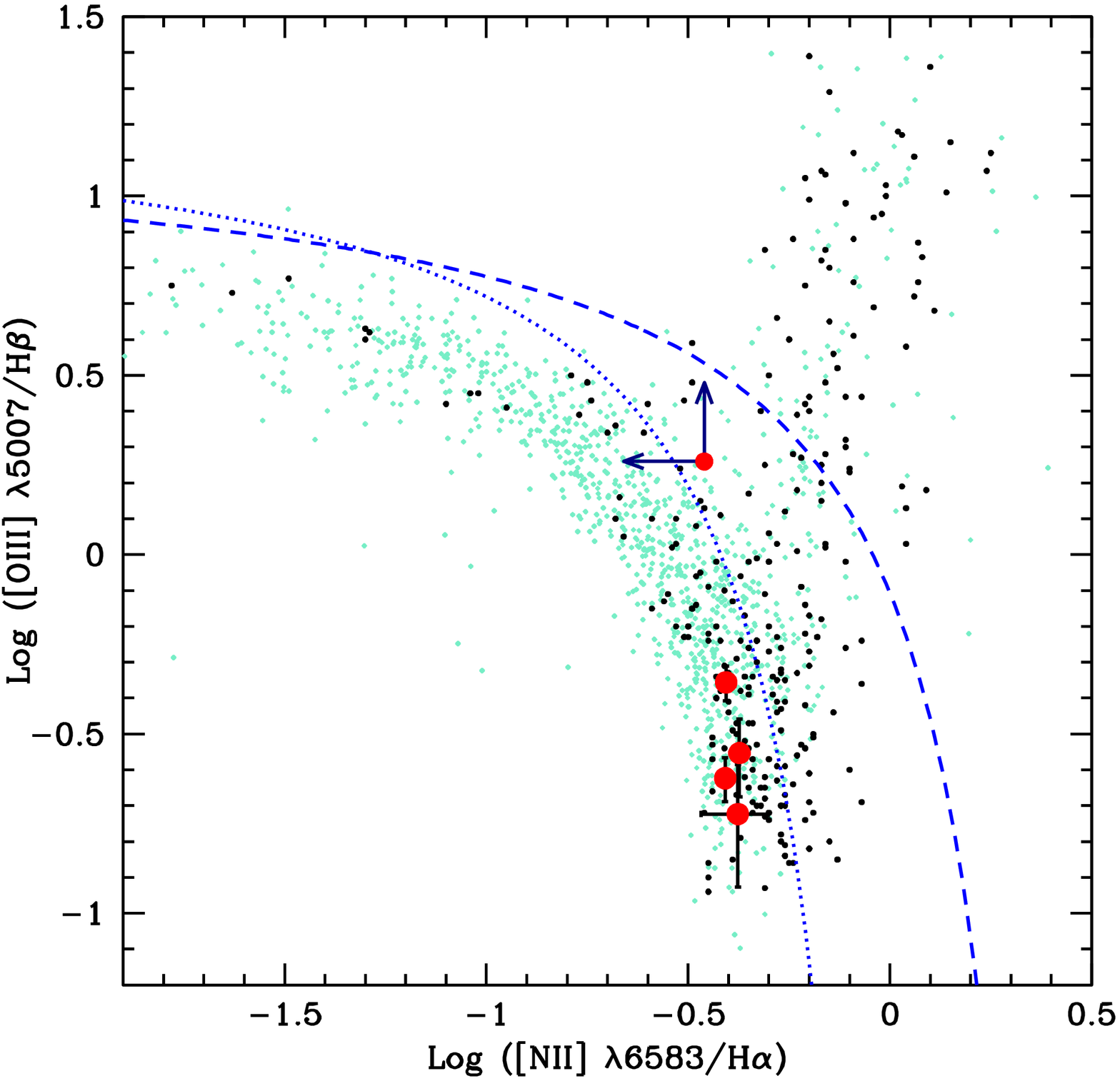}
\caption{[O\,{\sc iii}]/H$\beta$ vs. [N\,{\sc ii}]/H$\alpha$
diagnostic diagram. The galaxies considered in this work are shown
with large red circles. The small cyan circles are galaxies from the
KISS survey \citep{Salzer_etal_2005a} and the small black dots are
local starburst galaxies from \citet{Kewley_etal_2001b}. The dashed
line shows the locus of points which \citet{Kewley_etal_2001a}
consider to be the theoretical limit for starbursts, in the sense that
galaxies without an AGN component should fall below and to the left of
this line. The dotted line is an empirical determination by
\citet{Kauffmann_etal_2003a} of the same limit. It is evident from
their \NII/$\Ha$ and \OIII/$\Hb$ ratios that AGN contamination is
unlikely to be significant in our galaxies.}
\label{fig:AGN}
\end{figure}

Active Galactic Nuclei (AGN) contamination can be a significant
problem in galaxy emission line studies. In Fig.~\ref{fig:AGN} we plot
each galaxy in \NII/$\Ha$--\OIII/$\Hb$ space, the AGN diagnostic
developed by \citet*{Baldwin_etal_1981a} and
\citet{Veilleux_Osterbrock_1987a}. The harder radiation in AGN
compared to star-forming regions cause an enhanced \OIII/$\Hb$ ratio
with respect to \NII/$\Ha$. The loci derived by
\citet{Kewley_etal_2001a} and measured by \citet{Kauffmann_etal_2003a}
define the boundary beyond which AGN contamination is significant. AGN
contamination is therefore not significant for our galaxies.

\subsection{Dust extinction of emission line flux}\label{sec:extinction}
Dust extinction corrections are essential, particularly at shorter
wavelengths where attenuation due to dust is higher. This is
demonstrated in Section \ref{sec:CaII_SFRs}. For simplicity,
uncorrected flux values were used for those metallicity (and AGN)
indicators which are robust against relative differences in flux
calibration, otherwise extinction corrected flux values were used.

We estimate the galaxy extinction using the Balmer decrement between
$\Ha$ and $\Hb$. We assume an intrinsic Balmer decrement of 2.85 (for
case B recombination at $T=10^4\,\rm{K}$ and $n_e \sim 10^2 -
10^4\,\rm{cm}^{-3}$; \citealt{Osterbrock_1989}). This is
representative of \HII\ regions. The colour excess can then be defined
\begin{equation}
  E\left(B\!-\!V\right) \equiv
  \frac{-2.5\left[\log\left(\Ha/\Hb\right)_{\rm int}-\log\left(\Ha/\Hb\right)_{\rm obs}\right]}{k\left(\Hb\right)-k\left(\Ha\right)}\,,
\end{equation}
where $\left(\Ha/\Hb\right)_{\rm obs}$ is the Balmer decrement
observed after correction for stellar absorption,
$\left(\Ha/\Hb\right)_{\rm int}$ is the intrinsic Balmer decrement and
$k\left(\lambda\right)$ is an extinction curve. $k\left(\Ha\right)$
and $k\left(\Hb\right)$ are the values of $k\left(\lambda\right)$ at
6564 and $4862\,\rm{\AA}$, respectively from
\citet*{Cardelli_etal_1989a}. The extinction curve and this \EBV\ are
used to correct all the measured line using the following expression,
\begin{equation}
\log F_{\rm extncor}\left(\lambda\right) =
0.4k\left(\lambda\right)E\left(B\!-\!V\right)\log F_{\rm
  obs}\left(\lambda\right)\,,
\label{eqn:EBmV_cor}
\end{equation}
where $F_{\rm extncor}$ is the extinction corrected integrated line
flux and $F_{\rm obs}$ is the observed integrated line flux. In the
case of Balmer series lines, this observed flux has been corrected for
stellar absorption as discussed in Section
\ref{sec:general_anal}. Over the wavelength range we are interested in
the MW, Large Magellanic and Small Magellanic Clouds (LMC, SMC,
respectively) have similar extinction curves
(\citealt{Cardelli_etal_1989a};
\citealt*{Misselt_etal_1999a,Gordon_Clayton_1998a}), differing no more
than ten per cent at most. As is shown in Section
\ref{sec:caii_gen_results}, our galaxies more closely resemble the MW
than the LMC or SMC, hence we choose to use a MW attenuation curve for
the rest of our analysis. Aside from the choice of extinction curve,
the error in \EBV\ is dominated by the systematics involved in fitting
stellar absorption (see Section \ref{sec:general_anal}).

For J0912$+$5939 the $\Hb$ line is coincident with Galactic \NaI\
absorption while for J1219$-$0043 the $\Hb$ absorption line fit was
uncertain (see Section \ref{sec:general_anal}). Therefore, we do not
estimate \EBV\ for these lines of sight. As can be seen from
Table~\ref{tab:extinctions}, the derived extinctions for our sample
are above the average value observed in normal star-forming galaxies
at low redshift (\EBV~$\sim0.36\,\rm{mag}$), although they lie within
the distribution of values from such galaxies
\citep{Kennicutt_1992a,Moustakas_etal_2006a}.

\subsection{Systematic effects and error budget}\label{sec:systematics}
It is generally the case that systematic sources of error dominate the
error budget for both SFRs and metallicities derived empirically from
galaxy emission lines. As highlighted previously, the major sources of
systematic error are flux calibration (see Section
\ref{sec:reduction}), slit losses (see Section \ref{sec:geocor}),
fitting of stellar absorption along the Balmer series (see Section
\ref{sec:general_anal}) and extinction corrections (see Section
\ref{sec:extinction}). These are not independent of each other.
However, the most significant source of error was the fitting of the
stellar absorption. This corresponds to a maximum of $0.1\,\rm{dex}$
variation in metallicities and ten per cent variation in SFRs. The
effect on metallicity is small for two reasons.  Firstly, the
metallicity measurements rely on flux ratios of lines {\em close} in
wavelength (R23 metallicities being the most effected because of the
larger separation between \OII\ and \OIII).  Secondly, the empirically
calibrated metallicity indicators have only a shallow dependence on
the relevant line ratios. Due to the variation in calibration between
different metallicity indicators, the derived metallicities for any
single galaxy can vary by up to $0.3\,\rm{dex}$. This indicator
calibration variation therefore dominates the error in our derived
metallicities.

\section{General properties of our \CaII-selected galaxies}\label{sec:caii_gen_results}
From FORS2 acquisition and SDSS images we measured the galaxy
properties collected in Table~\ref{tab:galprop}. They include
measurements of their $r$--band magnitudes, rest-frame $B_j$ band
luminosities, QSO--galaxy impact parameters, morphological
classification (via spectral energy distribution (SED) fitting and
Hubble classification of acquisition images). We also measure velocity
dispersions ($\Ha$ line widths) from the galaxy spectra. It was only
possible to derive Hubble types for spatially resolved targets;
classification was aided by using {\sc galfit} \citep{Peng_etal_2002a}
to identify resolvable galaxy components.
\begin{table*}
  \begin{minipage}{0.8\textwidth}
    \caption{Measured SDSS $r$--band magnitudes, $B_j$ galaxy
    luminosities, QSO impact parameters, $\Ha$ line widths, best
    fitting CWW SEDs and morphological classifications.}
    \label{tab:galprop}
    \begin{tabular}{lcccccccc}
      \hline
      Object & $\zabs$ & $r/\rm{mag}$ & $L/L^*\,\left(B_j\right)$ & $b/\arcsec$ & $b/{\rm kpc}$ & $\sigma\left(\rm{H}\alpha\right)/\rm{km}\,\rm{s}^{-1}$ & SED & Hubble Type\\[0.5ex]
      \hline
      J0019$-$1053 & 0.347 & $20.21\pm0.03$   & $0.89 \pm 0.03$   & $3.5 \pm 0.2$  & $17 \pm 1$    & --          & Scd & --\\
      J0912$+$5939 & 0.212 & $21.2\pm0.4^\dagger$     & $0.13 \pm 0.05$   & $\lesssim 1.5$ & $\lesssim 5$  & $100 \pm 20$ & Im & --\\
      J1118$-$0021 & 0.132 &$17.22\pm0.01$ & $1.75 \pm 0.02$   & $9.9 \pm 0.2$  & $23.2 \pm 0.5$    & --          & Sbc & S0$_3$\\
      J1219$-$0043 & 0.448 & $21.2\pm0.1$     & $0.68 \pm 0.07$     & $1.3 \pm 0.2$  & $7 \pm 1$     & $100 \pm 40$ & Sbc & --\\
      J2246$+$1310 & 0.395 & $20.45\pm0.05$   & $1.01 \pm 0.05$   & $1.2 \pm 0.2$  & $6 \pm 1$     & $98 \pm 9$   & Scd & --\\
      \hline
    \end{tabular}

    \medskip    
    $^\dagger$This magnitude is derived by the deconvolution of PSF
    and Model magnitudes for the SDSS QSO.  As such, its error,
    derived from formal propagation of magnitude errors, is likely to
    be an underestimate and will be dominated by the uncertainty in
    the deconvolution.
  \end{minipage}
\end{table*}

\subsection{Galaxy luminosities}\label{sec:caii_gal_lum}
Most of our galaxies were classified photometrically in the SDSS and
therefore have SDSS $ugriz$ Model magnitudes and errors (the Model
profile is the De Vaucouleurs or exponential profile, whichever fit
has the lower $\chi^2$). J0912$+$5939 was too close to the QSO
sight-line for it to be detected independently in the SDSS. Instead we
calculated its magnitude by taking the difference between its QSO PSF
and Model magnitude. The assumption here is that the PSF profile will
predominantly consist of the QSO, whereas the Model magnitude will
consist of QSO + galaxy.  In practice some galaxy light will
contribute to the PSF magnitude as well, but this contamination is not
significant unless the galaxy is resolved
\citep[e.g.][]{Schneider_etal_2005}. Galaxy $r$--band magnitudes are
listed in Table~\ref{tab:galprop}.

Absolute $B_j$ magnitudes \citep{Maddox_Hewett_2006a} are derived from
$r$--band magnitudes, via the best fitting \citet*[][henceforth
CWW]{Coleman_etal_1980a} spectral energy distribution (SED). The best
fitting CWW SED is selected as that which has the smallest $\chi^2$
when fitted to the SDSS $ugriz$, extinction corrected, AB
magnitudes. The best fitting models are listed in
Table~\ref{tab:galprop}. The best-fitting SED is then used to
calculate the appropriate K-correction, $K\left(z\right)$, for the
$r$--band distance modulus, $m - M = 5\log\left(D_{\rm L}\right)-5 +
K\left(z\right)$, which results in an absolute $r$--band magnitude.
Here $D_{\rm L}$ is the luminosity distance to the galaxy in parsecs.
The $B_j-r\arcmin$ colour at $z=0$ for the best fit SED is then used
to convert to $B_j$ magnitudes. We note that the required
K-corrections are small ($\sim\!0.3\,\rm{mag}$) because the rest frame
$B_j$--band is similar to the SDSS $r$--band at the redshifts of our
galaxies. AB magnitudes are converted to standard Vega magnitudes
using the tabulation of \citet{Hewett_etal_2006a}, adopting a
$+0.03\,\rm{V\,mag}$ for Vega.  The absolute $B_j$ magnitude of an
$L^\ast$ galaxy in our cosmology is $M_{B_j}^\ast = -20.43\,\rm{mag}$
\citep{Norberg_etal_2002}. There is some evidence for an evolution of
$L^\ast$ with redshift, though the exact form of this evolution is
still under debate. We therefore choose to evolve $L^\ast$ with
redshift using the simple prescription of \citet{Ilbert_etal_2005a},
$M_{B_j}^\ast\!\left(z\right) = M_{B_j}^\ast\!\left(0\right) - z$,
which is reasonable at low-$z$ for the small redshift range we
sample. Choosing an evolving model for $L^\ast$ does not affect the
conclusions we draw from the measured luminosities. This leads to a
galaxy luminosity via the following simple relationship,
\begin{equation}
\log\left(L/L^\ast\right) = \log\left(F/F^\ast\right) = -0.4\left(M_{B_j}^{\rm gal}-M_{B_j}^\ast(z_{\rm gal})\right)\,.
\end{equation}
Formal errors on the SDSS magnitudes were propagated throughout. No
error from SED selection was folded into the analysis. The choice of
SED does not significantly effect the derived luminosities because the
variation in magnitude resulting from different SED fits is of order
the statistical error. No intrinsic error was assumed for the
AB-to-Vega magnitude conversion or for $M_{B_j}^\ast$. Note that most
of our galaxy have luminosities close to $L^\ast$ (see
Table~\ref{tab:galprop}). It is likely that we have biased our sample
to high luminosities by observing galaxies we could easily identify in
SDSS images (see Section \ref{sec:discussion}).

\begin{table*}
\begin{minipage}{1.0\textwidth}
\caption{SFRs and metallicities for DLA-selected galaxies at $z < 0.8$
  from the literature. Systematic errors are likely of the order
  $0.3\,\rm{dex}$ for metallicities and up to ten per cent for SFRs,
  respectively. See individual references for detailed discussion of
  the major sources of systematic error in their work.}
\label{tab:DLA_SFRs}
\begin{tabular}{lccccccccc}
\hline
&&& \multicolumn{2}{c}{\hspace{5ex}\vector(-1,0){20}\hspace{6ex} SFR / \Msunyr \vector(1,0){20}} & \multicolumn{4}{c}{\hspace{22ex}\vector(-1,0){80}\hspace{22ex}$12+\log\left(\rm{O}/\rm{H}\right)$\vector(1,0){80}}\\
\raisebox{1.0ex}{Object} & \raisebox{1.0ex}{$z$} & \raisebox{1.0ex}{\EBV$_{\rm gal}$} & \OII$_{\rm uncor}$ & $\Ha$ & N2 & O3N2 & R23 (lower) & R23 (upper) & \raisebox{1.0ex}{References}\\[1.0ex]
\hline
PKS 0439$-$433   & 0.101 & $0.211 \pm 0.007$ & $0.071 \pm 0.004$ & $0.55 \pm 0.01$ & $8.677 \pm 0.006$ & $8.766 \pm 0.006$ & $7.34 \pm 0.02$  & $9.011 \pm 0.006$ & 1\\
Q 0738$+$313     & 0.222 & --               & $<$0.003          & --             & --               & --               & --              & --               & 1\\
Q 0809$+$583     & 0.437 & --               & $0.46 \pm 0.04$   & --             & --               & --               & $7.23 \pm 0.06$  & $9.085 \pm 0.009$ & 1\\
                 &       & $0.49 \pm 0.03$   & $0.25 \pm 0.02$   & $5.1 \pm 0.5$   & $8.57 \pm 0.05$   & $8.73 \pm 0.04$   & $7.45 \pm 0.07$  & $8.99 \pm 0.02$   & 2\\
AO 0235$+$164    & 0.525 & --               & $1.89 \pm 0.05$   & --             & --               & --               & $8.06 \pm 0.05$  & $8.55 \pm 0.03$   & 1\\
B2 0827$+$293    & 0.526 & --               & $0.62 \pm 0.05$   & --             & --               & --               & $8.3 \pm 0.1$    & $8.4 \pm 0.1$     & 1\\
LBQS 0058$+$0155 & 0.612 & --               & $0.27 \pm 0.02$   & --             & --               & --               & --              & --               & 1\\
FBQS 0051$+$0041 & 0.740 & --               & 0.9               & --             & --               & --               & --              & --               & 3\\
\hline
\end{tabular}

\medskip
[1]~\citet{Chen_etal_2005a}. [2]~\citet{Gharanfoli_etal_2007a}. [3]~\citet{Lacy_etal_2003a}.
\end{minipage}
\end{table*}

\subsection{QSO--galaxy impact parameters}
For most objects, where both galaxy and QSO were identified as
distinct objects in the SDSS, we measure the QSO--galaxy impact
parameters by taking the difference between their J2000 co-ordinates,
assuming an error between co-ordinates of $0\farcs2$, the typical
accuracy of relative object positions in the SDSS. J0912$+$5939 is not
independently identified in the SDSS and we have no, higher
resolution, FORS2 images either. Given that we know the target is
relatively bright, based on galaxy luminosity and emission line
strength, we adopt a conservative upper limit to the impact parameter
of less than the SDSS fibre radius, $1\farcs5$.

\subsection{Galaxy velocity dispersions}
It is straight-forward to measure the line-widths, $\sigma_{\rm obs}$,
of our fitted $\Ha$ profiles for each galaxy. For $\sigma_{\rm obs} >
\sigma_{\rm ins}$, the instrumental profile width, direct application
of the convolution theorem, assuming Gaussian profiles, gives us a
reliable measure of the galaxy velocity dispersion,
\begin{equation}
  \sigma_{\rm gal}^2 = \sigma_{\rm obs}^2 - \sigma_{\rm ins}^2 -
  \sigma_{\rm th}^2\,,
\end{equation}
where $\sigma_{\rm th}$ is the thermal broadening of the $\Ha$ line
($= \sqrt{2kT/m} = 9\,\rm{km}\,\rm{s}^{-1}$ at $10^4\,\rm{K}$). The
instrumental profile widths, $\sigma_{\rm ins}$, for each setting were
measured from the extracted wavelength calibration arcs and the errors
were derived from the standard deviation from the mean of many arc
lines: $\sigma_{\rm 600B}= 1.9 \pm 0.1\,\rm{\AA}$, $\sigma_{\rm 300V}=
4.3 \pm 0.2\,\rm{\AA}$, $\sigma_{\rm 600RI}= 2.19 \pm 0.04\,\rm{\AA}$,
$\sigma_{\rm 600z}= 2.11 \pm 0.04\,\rm{\AA}$. For $\sigma \sim
\sigma_{\rm ins}$ the deconvolution is unlikely to be meaningful at
the S/N of our spectra, so we do not quote a result in this case. The
results are collected in Table~\ref{tab:galprop}.

To derive rotational velocities or dynamical masses from the measured
$\Ha$ line velocity dispersions requires many assumptions
\citep[e.g.][]{Erb_etal_2006b} and such an analysis is beyond the
scope of this paper. However, for galaxies at $z\sim0.25$
\citet{Rix_etal_1997a} show that on average, rotational velocity,
$v=1.67\sigma$. Though the dispersion in this relation is large it
shows that we expect our galaxies to have rotational velocities
$\sim\!170\,{\rm km}\,{\rm s}^{-1}$. For context, the rotational
velocity of the MW at $8\,\rm{kpc}$ is $v \simeq
220\,\rm{km}\,\rm{s}^{-1}$.

\section{Metallicities and star-formation rates for our galaxies}\label{sec:Z_SFR_caii}


We are principally interested in the emission-line metallicities and
SFRs of our observed galaxies. The wavelength coverage of the FORS2
(and SDSS) spectra allows comparison between three empirical
strong-line metallicity indicators: N2 ($\log\,$\NII/$\Ha$), O3N2
($\log\,$\OIII/$\Hb$ $-\log$\NII/$\Ha$) and R23
($\log\,($\OII$+$\OIII$) - \log\,\Hb$). R23 requires a dust extinction
correction and these were calculated using the Balmer decrement (see
Section \ref{sec:extinction}).  Dust extinction corrections are also
essential for realistic SFR measurements (see Section
\ref{sec:dla_results} and Section \ref{sec:discussion}).  As mentioned
in Section \ref{sec:systematics}, the order of magnitude of these
effects is $\approx 0.1\,{\rm dex}$ on metallicities and up to ten per
cent on SFRs. For this work we consider SFRs based on $\Ha$ and \OII\
because, whilst direct measurements from $\Ha$ are more reliable, the
\OII\ indicator is useful when comparing these results with others in
the literature (see Section \ref{sec:dla_results}).

Prior to presenting our new results, we review the results for
DLA-selected galaxies already in the literature. Thus providing a
background on which to base later conclusions.

\subsection{Metallicities and star-formation rates of DLA-selected galaxies}\label{sec:dla_results}

In the literature to date there have been just eight measurements of
emission lines in DLA-selected galaxies at $z < 0.8$ from which
classical strong line metallicity and SFR indicators can be derived
\citep{Lacy_etal_2003a,Chen_etal_2005a,Gharanfoli_etal_2007a}. This is
in no small part due to the difficulty in identifying such DLAs
without a UV space spectrograph. Only two of these eight galaxies have
$\Ha$ line flux measurements; SFRs have been largely based on \OII\
and it has not been possible to estimate dust extinction in these
galaxies. As we show in Section \ref{sec:CaII_SFRs}, correcting for
dust extinction is vital for obtaining reliable SFR measurements.
Here, beginning with the measured integrated line fluxes, we re-derive
these SFRs and metallicities using exactly the same assumptions that
have been used to derive our \CaII-selected galaxy SFRs so that a fair
comparison can be made between them. The results are shown in
Table~\ref{tab:DLA_SFRs}.

R23 is sensitive to reddening so the only reliable metallicities are
those for PKS 0439$-$433 and Q 0809$+$583 where $\Ha$ is detected and
an extinction correction can be made. For these objects N2 and O3N2
show that we should consider the upper branch of R23.

\subsection{Metallicities of our \texorpdfstring{\CaII}{CaII}-selected galaxies}\label{sec:caii_Z}
There are currently no metallicity measurements of \CaII\ absorbers in
either emission or absorption. Here we use the metallicity indicators
N2(linear) and O3N2, as calibrated by \citet[][henceforth
\citetalias{Pettini_Pagel_2004a}]{Pettini_Pagel_2004a}, and R23, as
calibrated by \citet*{Kobulnicky_1999a}\footnote{Whilst more modern
calibrations exist, they also include more parameters, hence we
utilize this R23 calibration. It will make, at most, a $0.1\,\rm{dex}$
difference to the R23 results.}, to provide the first emission-line
metallicity measurements.  Where there was no line detection for \NII,
\OII\ or \OIII\ lines, 3$\sigma$ upper limits were used. J0912$+$5939
does not have an $\Hb$ detection, as previously described (see Section
\ref{sec:extinction}). Therefore, we do not give results for
metallicities involving $\Hb$ in this case. The results are collected
in Table~\ref{tab:CaII_Z}.

\begin{table}
\setlength{\tabcolsep}{4.5pt}
\caption{\CaII\ absorption-selected galaxy emission-line metallicities
for various strong line indicators. The limit for J0912$+$5939 is
derived without correcting for \EBV\ or stellar absorption at
$\Ha$. The systematic errors in each metallicity measurement are $\sim
0.3\,\rm{dex}$.}
\label{tab:CaII_Z}
\begin{tabular}{lcccc}
  \hline
  &\multicolumn{4}{c}{$12+\log\left(\rm{O}/\rm{H}\right)$}\\[0.8ex]
  \raisebox{1.0ex}{Object} & N2(linear) & O3N2 & R23(lower) & R23(upper)\\[0.0ex]
  \hline
  J0019$-$1053  &  $8.67 \pm 0.01$ & $8.80 \pm 0.02$ & $7.7 \pm 0.1$ & $8.94 \pm 0.04$\\
  J0912$+$5939  &  $<8.6$  & -- & -- & --\\
  J1118$-$0021  &  $8.687 \pm 0.007$ & $8.79 \pm 0.03$ & $7.8 \pm 0.1$ & $8.88 \pm 0.04$\\
  J1219$-$0043$^\dagger$  &  $8.69 \pm 0.05$ & $8.84 \pm 0.06$ & $7.45 \pm 0.07$ & $9.01 \pm 0.02$\\
  J2246$+$1310  &  $8.67 \pm 0.01$ & $8.71 \pm 0.02$ & $7.79 \pm 0.05$ & $8.85 \pm 0.02$\\
  \hline
\end{tabular}

\medskip
$^\dagger$Metallicity measurements for this object are strictly lower
limits because no correction for \EBV\ was made.
\end{table}

All the measured metallicities are around or above the solar value of
$8.66 \pm 0.05$ \citep{Asplund_etal_2005a}. In this regime the strong
line metallicity indicators are poorly calibrated despite extensive
studies \citep[e.g.][]{Bresolin_2006a} due the small number of known
super-solar metallicity galaxies which are bright enough to study. For
this reason we utilize only the linear form of the N2 index, as laid
out by \citetalias{Pettini_Pagel_2004a}, as the cubic form has an
upturn precisely in this uncalibrated metallicity regime. The N2 index
is known to saturate around solar metallicity -- there are indications
of this in our data when comparing it to the O3N2 and R23 indices. The
N2 and O3N2 values indicate that we should consider the upper branch
R23 data. R23 upper branch metallicities above solar are only
calibrated using photoionization models because metal-line cooling
makes recombination lines too weak to detect in this regime for
empirical calibration. All our metallicities agree within
$0.3\,\rm{dex}$; flux calibration errors are therefore not likely to
dominate the results and, in any case, the indicators were designed to
be robust against such errors (Section \ref{sec:relfluxcal}). It is,
in fact, the $0.3\,\rm{dex}$ variation between different indicators,
due to {\em their} calibration which will dominate our metallicity
errors. We can, nevertheless, conclude that these galaxies have
near-solar metallicities.

\begin{figure}
  \includegraphics[origin=c,scale=0.40,trim=0 0 0 0,angle=0,clip=true]{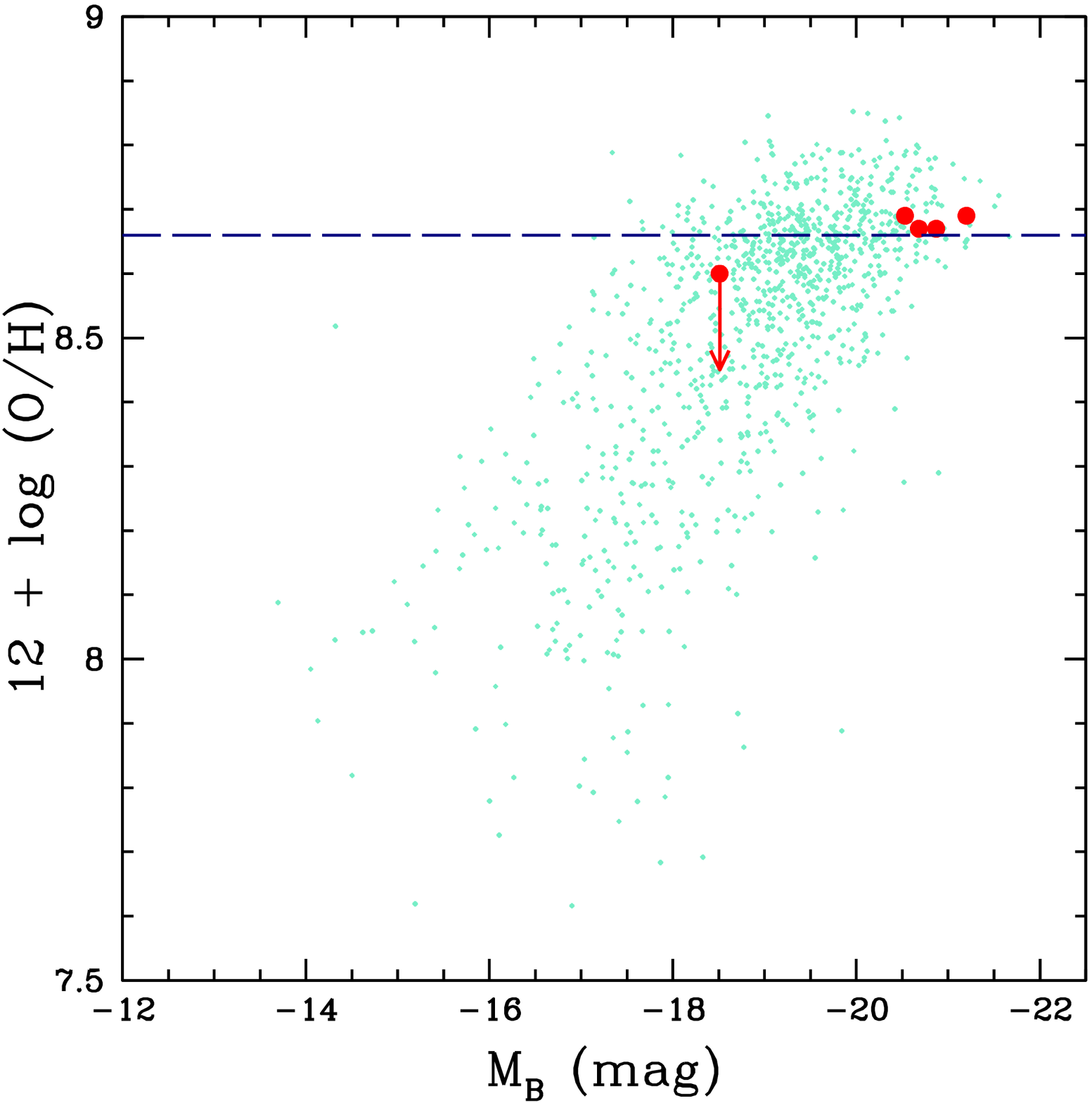}
\caption{A plot of metallicity, based on the N2 index, vs. luminosity
(absolute $B$ band magnitude). In cyan are the results for local
galaxies from the KISS survey \citep{Salzer_etal_2005a}. In red are
the \CaII-selected galaxies from this paper. An $L^\ast$ galaxy at
$z=0$ has $M_{B_j}=-20.43\,\rm{mag}$ \citep{Norberg_etal_2002}; a
$0.05L^\ast$ galaxy has $M_{B_j}=-17.17\,\rm{mag}$. $L^\ast$ at the
redshift of our galaxy sample is $\sim\!0.3\,\rm{mag}$ brighter in the
model we adopt. The dashed line indicates solar metallicity
\citep{Asplund_etal_2005a}. The \CaII-selected galaxies generally lie
near the extreme of the distribution for local galaxies.}
\label{fig:Z-L}
\end{figure}

It is not currently possible to compare the emission line
metallicities with absorption line metallicities from strong \CaII\
systems because there are none in the literature and new absorption
measurements would require a space-bourne UV spectrograph.  However,
if the \CaII\ absorber population overlap with the DLA population, it
is interesting that these emission-line metallicities are so high
compared to the generally low absorption-line metallicities measured
in DLAs, i.e. typically $-0.5\,\rm{dex}$ at $\zabs \sim 0.4$. The
galaxies presented here lie on the galaxy metallicity-luminosity
relationship for local galaxies from the KISS survey
\citep{Salzer_etal_2005a}, though they are at the upper end of the
luminosity range, as shown in Fig.~\ref{fig:Z-L}. Previous studies of
absorption-selected galaxy emission-lines have found similarly high
metallicities (e.g. Table \ref{tab:DLA_SFRs};
\citealt*{Ellison_etal_2005a}), so our new results are perhaps not
surprising and confirm that absorption-selected galaxies of all
flavours {\em can} be metal-rich.

\subsection{Star-formation rates of our \texorpdfstring{\CaII}{CaII}-selected galaxies}\label{sec:CaII_SFRs}

We consider SFRs based on $\Ha$, calibrated by
\citet{Kennicutt_1998a}, and \OII\, calibrated by
\citet*{Kewley_etal_2004}. These are both based on a
\citet{Salpeter_1955a} $0.1 - 100\,\rm{M}_\odot$ initial mass
function, giving
\begin{equation}
\rm{SFR}\left(\rm{H}\alpha\right) = 7.9\times10^{-42}\times
L\left(\rm{H}\alpha,\,\rm{erg}\,\rm{s}^{-1}\right)\,\rm{M}_\odot\,\rm{yr}^{-1}\,,
\end{equation}
\begin{equation}
\rm{SFR}\left(\left[\rm{O}\,\textrm{\sc ii}\right]\right) =
6.58\times10^{-42}\times L\left(\left[\rm{O}\,\textrm{\sc
ii}\right],\,\rm{erg}\,\rm{s}^{-1}\right)\,\rm{M}_\odot\,\rm{yr}^{-1}\,.
\end{equation}
The measured line fluxes are corrected for extinction and converted to
line luminosities based on their redshifts. No account is taken of the
dispersion in the above luminosity--SFR relationships when calculating
the formal errors. The measured luminosities and SFRs are in
Table~\ref{tab:CaII_SFRs}. We also include an \OII\ SFR measurement
which has {\em not} been corrected for extinction for comparison with
other results in the literature in Section \ref{sec:dla_results}.
Aperture corrections are calculated as described in Section
\ref{sec:geocor} and the resulting corrections and SFRs are shown in
Table~\ref{tab:geocor}. The systematic difference observed between
SFRs based on \OII\ and $\Ha$ is due to their calibration. Based on
its SDSS spectrum \citet{Brinchmann_etal_2004a} derive a SFR of
$7\pm2\,\Msunyr$ for J1118$-$0021, which is consistent with our
derivation of $6.2\pm0.7\,\Msunyr$ from a FORS2 spectrum.

\begin{table}
\setlength{\tabcolsep}{2.0pt}
\caption{\CaII\ absorption-selected galaxy SFRs and line
  luminosities. Systematic errors are up to ten per cent.}
\label{tab:CaII_SFRs}
\begin{tabular}{lccccc}
  \hline
  &\multicolumn{2}{c}{$L$ / $\times10^{40}\,\rm{erg}\,\rm{s}^{-1}$}&\multicolumn{3}{c}{$\rm{SFR}$ / $\rm{M}_\odot\,\rm{yr}^{-1}$}\\
  \raisebox{1.0ex}{Object} & \OII & $\Ha$ & \OII$_{\rm uncor}$ & \OII & $\Ha$\\[1.0ex]
  \hline
  J0019$-$1053 & $210 \pm 20$ & $330 \pm 40$ & $1.37 \pm 0.06$ & $14 \pm 2$ & $26 \pm 3$\\
  J0912$+$5939 & $3.5 \pm 0.7^\dagger$ & $4.5 \pm 0.7^\dagger$ & $0.23 \pm 0.04$ & -- & $0.35 \pm 0.06^\ddagger$\\
  J1118$-$0021 & $29 \pm 3$ & $35 \pm 3$ & $0.14 \pm 0.01$ & $1.9 \pm 0.2$ & $2.8 \pm 0.3$\\
  J1219$-$0043 & $8.2 \pm 0.6^\dagger$ & $24 \pm 2^\dagger$ & $0.54 \pm 0.04$ & -- & $1.9 \pm 0.1^\ddagger$\\
  J2246$+$1310 & $77 \pm 4$ & $91 \pm 4$ & $0.90 \pm 0.03$ & $5.1 \pm 0.3$ & $7.2 \pm 0.3$\\
  \hline
\end{tabular}

\medskip
$^\dagger$This luminosity is {\em not} corrected for dust extinction
  due to the absorbing galaxy.

$^\ddagger$Lower limit to SFR($\Ha$) based on uncorrected line flux.
\end{table}

\begin{table}
\caption{Geometric slit corrections and corrected \CaII\
absorption-selected galaxy SFRs. Systematic errors are up to ten per
cent. A correction was not possible for J0912$+$5939, for which only a
SDSS galaxy spectrum is available, but it is unlikely to be
significant (see Section \ref{sec:caii_gal_lum}). These results
constitute our best estimates of the SFRs in these galaxies.}
\label{tab:geocor}
\begin{tabular}{lccc}
  \hline
  & geometric &\multicolumn{2}{c}{$\rm{SFR}$ / $\rm{M}_\odot\,\rm{yr}^{-1}$}\\
  Object & correction & \OII & $\Ha$\\[0.5ex]
  \hline
  J0019$-$1053 & 1.11 & $16\pm2$ & $29\pm3$\\
  J1118$-$0021 & 2.20 & $4.2\pm0.4$ & $6.2\pm0.7$\\
  J1219$-$0043 & 1.41 & -- & $2.7\pm0.1^\dagger$\\
  J2246$+$1310 & 1.50 & $7.7\pm0.5$ & $10.8\pm0.5$\\
  \hline
\end{tabular}

\medskip
$^\dagger$Lower limit to SFR($\Ha$) based on uncorrected line flux.
\end{table}

\begin{figure}
  \includegraphics[origin=c,width=\columnwidth,trim=0 0 0 0,angle=0,clip=true]{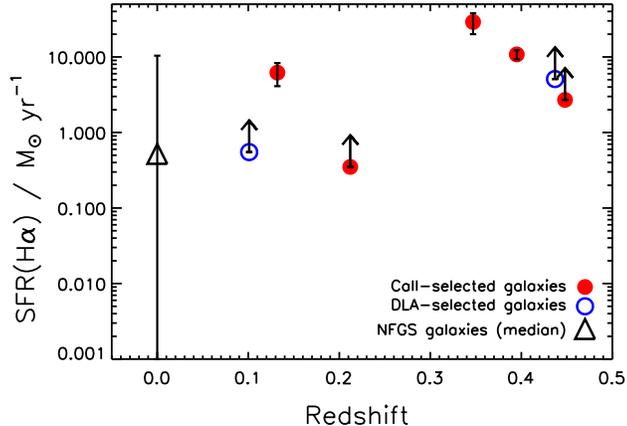}
\caption{SFR($\Ha$) (corrected for extinction and slit loss)
vs. redshift for all the absorbers with $\Ha$ line coverage.  Solid
red circles are \CaII\ absorbers, hollow blue circles are DLAs. For
clarity, errors are 3$\sigma$. DLAs were not corrected for slit loss
and hence SFRs are lower limits. The black triangle at $z=0$, is the
median SFR($\Ha$), in the Nearby Field Galaxy Survey, from
\citet{Kewley_etal_2002a}. The error bar represents the 1$\sigma$
standard deviation from the median for the sample.}
\label{fig:SFR_Ha}
\end{figure}

\begin{figure}
  \includegraphics[origin=c,width=\columnwidth,trim=0 0 0 0,angle=0,clip=true]{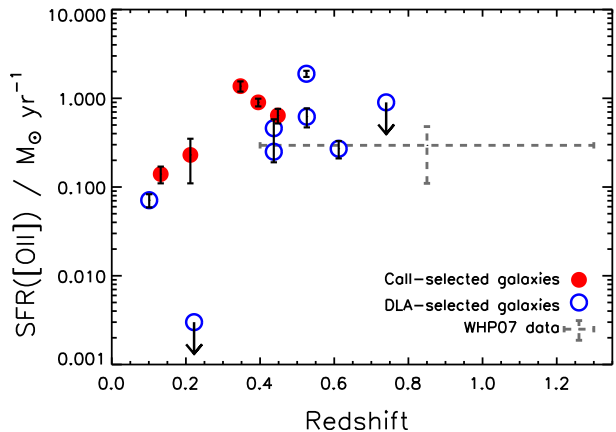}
\caption{SFR(\OII) (uncorrected for extinction or slit loss)
vs. redshift for all the absorbers. Solid red circles are \CaII\
absorbers, hollow blue circles are DLAs. For clarity, errors are
3$\sigma$. The two measurements at $z=0.437$ are independent
measurements of the same galaxy as indicated by the joint error
bar. The dashed line data are the higher-$z$ SFR range from stacking
\CaII\ absorbers derived by \citetalias{Wild_etal_2007a}. }
\label{fig:SFR_OII}
\end{figure}

The results based on $\Ha$ seem to suggest we are selecting actively
star-forming galaxies. The median SFR for local galaxies, taken from
the NFGS, is $0.5\,\rm{M}_\odot\,\rm{yr}^{-1}$
\citep{Kewley_etal_2002a}, while the median SFR in our galaxies is
$10.8\,\rm{M}_\odot\,\rm{yr}^{-1}$ (or $6.2\,\rm{M}_\odot\,\rm{yr}^{-1}$
when lower limits are included). Fig.~\ref{fig:SFR_Ha} shows the $\Ha$
SFRs measured in our \CaII-selected galaxies, DLA-selected galaxies
and the local NFGS median. We emphasize that the same conclusion would
not be reached from an analysis based purely on direct flux
measurements of \OII; this would give a mean SFR of
$0.6\,\rm{M}_\odot\,\rm{yr}^{-1}$.  Fig.~\ref{fig:SFR_OII} shows the
\OII\ SFRs measured in our \CaII-selected galaxies. The same figure
also shows the SFRs from the DLA-selected galaxies in Table
\ref{tab:DLA_SFRs} and the higher redshift \CaII\ absorber sample mean
SFR measured by \citetalias{Wild_etal_2007a}. Given that our data has
an inherent luminosity bias (see Section \ref{sec:caii_gal_lum}), our
SFRs are in good agreement with the higher redshift work of
\citetalias{Wild_etal_2007a}, whose sample does not suffer from the
same bias.

\section{Discussion}\label{sec:discussion}

The most important results from this work are the galaxy emission
line metallicities and SFRs. Before discussing these we briefly review
what we have learnt about the general properties of our sample to set
the context in which to consider the main results.

\subsection{General properties of our \texorpdfstring{\CaII}{CaII}-selected galaxies}

The dominant property of the galaxies in our sample, with reference to
the derived metallicities and SFRs, are their $\sim L^\ast$
luminosities. It is likely this is a selection bias, which is
highlighted by the fact that the lowest luminosity galaxy is one we
did not identify through SDSS imaging, but rather through its galaxy
spectrum superimposed on the QSO spectrum.

Given their luminosities, one might expect these galaxies to have
rotational speeds similar to the MW. The measured $\Ha$ line widths of
$\sigma \approx 100\,{\rm km\,s}^{-1}$ provide lower limits which are
not inconsistent with this. However, given a random distribution of
galaxy inclinations one might also expect some of these lower limits
to be nearer the MW rotational speed if the $\Ha$ emission is truly
tracing galactic rotation. Focusing on their velocity structures using
with long-slit or integral field spectra would demonstrate whether
this is the case or whether the $\Ha$ emission traces other processes
like outflows from the \HII\ regions.

The QSO--galaxy impact parameters in Table~\ref{tab:galprop} are
similar to those of DLAs \citep[e.g.][]{Steidel_1993a} which is
consistent with the \CaII\ absorber population significantly
overlapping with that of DLAs \citepalias{Wild_etal_2006a}. Whilst
this somewhat justifies our comparison between DLA and \CaII\ absorber
SFRs, it is based on small number statistics. One might generally
expect there to be a greater distribution of \CaII\ absorber impact
parameters; DLAs and other absorbers, such as those with strong \MgII,
certainly have a broader impact parameter distribution. Recent
$K$--band imaging of \CaII\ absorbers by \citet{Hewett_Wild_2007b}
confirms this at high redshift, where they find mean impact parameters
of $24\,\rm{kpc}$. However, it is not obvious how our selection
procedure would bias the data in this respect.

The SED fits to the sample suggest that the \CaII-selected galaxies
are normal star-forming spirals. Since spirals should have relatively
high gas cross-sections, it is not surprising that such galaxies are
selected via their absorption properties. However, in general, direct
imaging of low-redshift DLAs reveals a mix of host galaxy types,
including irregular, spiral and low-surface brightness galaxies
\citep[e.g.][]{LeBrun_etal_1997a,Rao_etal_2003a}. This is supported by
a blind 21-cm emission study of local galaxies
\citep*{Ryan-Weber_etal_2003a}. It is unclear whether the predominance
of spirals in our \CaII-selected sample is a consequence of the \CaII\
selection itself or our aforementioned luminosity bias. More detailed
imaging and spectroscopic study of velocity structure of a larger
sample is required to make more specific conclusions about the general
properties of these \CaII-selected galaxies and to avoid the
luminosity bias inherent in our current sample. Nevertheless, some
interesting conclusions can be drawn from the metallicities and SFRs
measured in our galaxies.

\subsection{Metals in \texorpdfstring{\CaII}{CaII}-selected galaxies}

At least four out of the five galaxies studied are metal-rich, with
solar or super-solar oxygen abundances (Table \ref{tab:CaII_Z}).
Interestingly, similar values have been found in both cases where
emission line abundances could be deduced in the host galaxies of
known DLAs at $z < 0.8$ (Table \ref{tab:DLA_SFRs}). Potentially, the
oxygen abundance could be significantly super-solar in the galaxies
where we estimate (O/H)\,$\simeq$(O/H)$_{\odot}$, because the
strong-line indices at our disposal are poorly calibrated in the
super-solar regime and the whole issue of how best to measure
metallicities in nebulae with (O/H) $>$ (O/H)$_{\odot}$ remains
controversial \citep[e.g][]{Bresolin_2006a,Peimbert_etal_2006a}.

Such high metallicities contrast with the generally sub-solar element
abundances measured in most DLAs from interstellar absorption lines.
In the latest census by \citet{Kulkarni_etal_2006b}, the average
(column density weighted) metallicity of confirmed DLAs is $\langle
Z_{\rm DLA} \rangle \simeq 1/7\,Z_{\odot}$ at $\langle z \rangle
\simeq 0.5$. There are several reasons which could explain such a
dichotomy.

One possibility is that such differences arise from the fact that
different elements are being considered in the two sets of
measurements. While emission line abundances measure (O/H), the
interstellar determinations of $Z_{\rm DLA}$ are based on Zn which
behaves like a Fe-peak element \citep*[e.g.][]{Chen_etal_2004a}.
However, in most DLAs -- and indeed in Galactic stars -- the
$\alpha$-element enhancement expected at metallicities
[Fe/H]$\,=\,-0.82$ (i.e.~1/7 solar) is less than a factor of three.
Thus, this possibility is unlikely to explain fully the difference
between our observations here and the typical DLA absorption line
abundances.

A second option, which has already been mentioned, is the luminosity
bias of our selection. As we favoured brighter galaxies in this
initial study, we are also likely to have selected the more metal-rich
galaxies within the presumably broad distribution of metallicities of
DLA hosts (given the wide spread of values of [Zn/H] measured in
absorption).

Thirdly, it is conceivable that in some cases we may have
misidentified the absorbing galaxy and that the host galaxy of the DLA
is a low luminosity dwarf below the detection limit of our images
located close to the bright galaxy we have identified. Galaxy
clustering certainly makes this a possibility in some instances,
although it is unlikely that we should be misled in every case. The
prevalence of this scenario is being tested by searches for galaxies
near faded gamma-ray bursts
\citep[e.g.][]{Jakobsson_etal_2004a,Ellison_etal_2006a} and for DLA
galaxy hosts along sight-lines where high-$z$ Lyman-limit systems
block the QSO flux from blue imaging bands \citep*{OMeara_etal_2006a}.

A fourth possibility, and potentially the most interesting to explore
further, is that there may be systematic differences between emission-
and absorption-based abundances if the two sets of data sample
different regions within the galaxies.  One could envisage a scenario
where the nebular emission lines are stronger in the inner, higher
surface brightness, regions, while the cross-section of the neutral
gas producing the DLA is larger in the outer parts. In the presence of
radial abundance gradients, a systematic offset of the sign and
magnitude observed could result. This possibility has already been
discussed by \citet{Chen_etal_2005a} and \citet{Bowen_etal_2005a} and,
while certainly plausible, its importance has yet to be quantified.
What is required is a direct comparison of emission- and
absorption-based abundances in the same galaxies and using elements
which share the same stellar nucleosynthesis and degree of
interstellar depletion. These requirements have so far been met in
only one case, studied by \citet{Bowen_etal_2005a}, where no offset
was found between nebular and DLA abundances. However, the galaxy in
question is a dwarf spiral where no abundance gradient was expected
over the radial distance probed.

In this respect, the sample of galaxies presented here, while still
small, constitutes a prime set of targets for future ultraviolet
observations once the Cosmic Origins spectrograph (COS) in installed
on a refurbished \emph{HST}. In particular, by measuring the abundance
of sulphur (an undepleted, $\alpha$-capture element) in the \CaII\
absorption systems, it will be possible to (i) obtain a quantitative
estimate of radial abundance gradients in galaxies at earlier epochs
-- a parameter which is a fundamental importance in galactic chemical
evolution models \citep[e.g.][]{Carigi_etal_2005a} and (ii) clarify
the reasons why $Z_{\rm DLA}$ remains sub-solar at all redshifts.

Recently, \citet{Herbert-Fort_etal_2006a} studied `metal strong' DLAs
(MSDLAs), which are defined to have $\log\left(N_{\rm
Zn\textsc{ii}}\right) \ge 13.15$ or $\log\left(N_{\rm
Si\textsc{ii}}\right) \ge 15.95$, where the column densities are
measured in $\rm{atoms}\,\rm{cm}^{-2}$. Given the high metallicities
of our \CaII-selected galaxies, it could be that the \CaII\ absorber
population overlaps more strongly with that of MSDLAs compared to that
of DLAs in general. Studying the \ZnII\ and \SiII\ line strengths in a
sample of \CaII\ absorbers would be interesting in this respect. This
is already possible at absorption redshifts
$0.65\lesssim\zabs\lesssim1.3$ since the \CaII\ and \ZnII\ transitions
fall in the optical region. We have begun such a study with high
resolution QSO spectra from the VLT. Clearly, for the redshifts
studied here ($\zabs\lesssim0.5$), UV spectra are required.

\subsection{Star formation in \texorpdfstring{\CaII}{CaII}-selected galaxies}\label{sec:SFR_discussion}
The galaxies in our sample are actively star-forming. Luminosity bias
probably affects the measured SFRs more than the measured
metallicities because the observed luminosity is, in part, caused by
current star formation, while the high metallicities more strongly
reflect past star formation. Thus, although it is possible that most
low-$z$ \CaII-selected galaxies could have near solar metallicities,
it is unlikely they all have such high SFRs. We reiterate that the
systematic effects discussed in Section \ref{sec:systematics} will
have reduced the SFRs we measure and so all our SFRs are technically
lower limits.

It is clear from Fig.~\ref{fig:SFR_OII} that the DLA-selected galaxies
in the literature have similar \OII\ emission-line SFRs, before
correcting for extinction, to our \CaII-selected galaxies. This is
unsurprising, given that they were generally selected in a similar way
to how we chose the \CaII\ galaxy sample (i.e.~relatively bright
galaxies were found close to the QSO sight-lines). However, what {\em
is} surprising is that this suggests that the DLA-selected galaxies
should have similar {\em extinction corrected} SFRs to our \CaII\
galaxies (assuming the star formation regions in those galaxies have
similar levels of dust extinction); i.e.~the DLA galaxies are also
actively star-forming. The amount of star-formation associated with
these DLAs has therefore been underestimated in the
literature. Extinction corrections are important in star-forming
regions, irrespective of the amount of dust associated with the
absorber itself, which is generally very low (e.g.
\citealt*{Ellison_etal_2005b}; \citealt{Murphy_Liske_2004a};
\citetalias{Wild_etal_2006a}). Previous authors have generally found
much lower rates of star-formation associated with DLAs (e.g.
\citealt{Wolfe_Chen_2006a}; \citealt{Hopkins_etal_2005a};
\citealt{Zwaan_etal_2005a,Wolfe_etal_2003a}).  This is perhaps an
indication that DLAs, whilst not being the sites of {\em in situ}
star-formation, are nonetheless closely associated with it.

If one naively attempts to directly compare our sample's \OII\ SFRs to
the higher redshift measurement of \citetalias{Wild_etal_2007a}, who
stack many \CaII\ absorption spectra to measure an SFR based on \OII,
one concludes that our SFRs are too high to be applicable to the
general \CaII-selected galaxy population. This is most likely due to
differences between the impact parameter and luminosity distributions
of the two samples. For example, two broad scenarios could plausibly
explain the difference between the relatively unbiased
\citetalias{Wild_etal_2007a} results and our own:

\begin{enumerate}
\item The relatively low \OII\ star-formation signal measured in
  \citetalias{Wild_etal_2007a} could be dominated by $\sim\!L^\ast$
  galaxies at low impact parameters (i.e.~inside the SDSS fibre
  radius). These galaxies dominate the evolution in the Madau
  diagram. That is, only some \CaII\ absorbers in this scenario would
  be closely associated with the sites of star-formation. Assuming
  this to be true, the following simple correction to the
  \citetalias{Wild_etal_2007a} results allows a comparison with our
  more directly measured SFRs, thereby providing a crude estimate of
  the luminosity bias in our sample. Firstly, at high redshift
  \citet{Hewett_Wild_2007b} find that only thirty per cent of the
  excess galaxy luminosity associated with the \CaII\ absorbers falls
  within the SDSS fibres, thus the \citetalias{Wild_etal_2007a}
  results must be altered to account for this. Secondly,
  \citetalias{Wild_etal_2007a} corrected their SFR measurement only
  for the very small dust extinction found in the absorbers
  themselves, not that appropriate to $\sim\!L^\ast$ galaxies. For
  local galaxies, the average $\EBV$ is $0.36$\,mag
  \citep{Kennicutt_1992a}, corresponding to an extinction factor of
  $4.54$ at \OII. Thus, the \citetalias{Wild_etal_2007a} SFR of
  $0.2\,\Msunyr$ translates to $100/30 \times 4.54 \times 0.2 =
  3.0\,\Msunyr$ in this scenario. This is $0.32$ times the mean
  dust-corrected SFR in our sample ($9.3\,\Msunyr$). From the Madau
  diagram \citep{Hopkins_2004a}, we expect a factor of $\sim 3.5$ drop
  in SFR density between the mean redshifts of the
  \citetalias{Wild_etal_2007a} sample ($\overline{z}=0.850$) and ours
  ($\overline{z}=0.285$). That is, our galaxy sample would appear to
  be $\sim3.5/0.32 = 11$ times more star-forming than expected for
  \CaII-selected galaxies at low-$z$.

\item Star-formation could be ubiquitous amongst \CaII\ absorbers. In
  this case its signature would be dominated by low level {\em in
  situ} star-formation and the required reddening corrections would be
  low, i.e.~those used by \citetalias{Wild_etal_2007a}. If this
  scenario were true then we have no basis on which to compare our low
  redshift results to this higher redshift one. However, our
  luminosity bias would be significant since our galaxy sample would
  represent highly unusual \CaII\ galaxies.
\end{enumerate}

Knowing \EBV\ from the galaxy emission lines in the
\citetalias{Wild_etal_2007a} sample would help distinguish between
these scenarios. Unfortunately, their stacked \CaII\ absorber spectrum
has too low S/N to put meaningful constraints on \EBV\ via $\Hb$ and
it is not clear that the $\Hb$ limit from their stacked \MgII\ DLA
sample applies equally to the \CaII\ sample.  Nevertheless, the limits
on \EBV\ from their \MgII\ sample favour scenario (ii) above. Thus,
while \citetalias{Wild_etal_2007a} favour low levels of {\em in situ}
star-formation associated with \CaII\ absorbers and DLAs, our work
provides clear examples of highly star-forming \CaII\ galaxies. It
seems likely that some mix of these scenarios more closely reflects
reality. Indeed, \citet{Hewett_Wild_2007b} find that their higher
redshift \CaII\ absorbers have a stronger galaxy
luminosity--dependence on the absorber cross-section ($\sigma \propto
L^{0.7}$) compared with that for other QALs, such as \MgII\ absorbers
($\sigma \propto L^{0.4}$). A study of the SFRs in less luminous
\CaII-selected galaxies will help to elucidate the situation; we are
currently undertaking such a study on the Gemini telescope.

\section{Conclusions}\label{sec:conclusions}

We have presented the first direct study of \CaII-selected galaxies.
The five galaxies in our sample are metal rich and actively
star-forming. We have also demonstrated the importance of both stellar
absorption along the Balmer series and dust extinction in such galaxy
studies. Prior to this work, the literature contained emission-line
fluxes of just seven DLA-selected galaxies at $z<0.8$ from which SFRs
could be derived. In only two of these galaxies was it possible to
make the all-important dust corrections required for a robust
interpretation of the observed SFRs. Given the probable overlap
between the \CaII\ and DLA populations, the current work represents a
significant contribution to our fledgling understanding of low-$z$
DLA-selected galaxies and the galaxy--absorber connection.

It is likely that luminosity bias, introduced during our sample
selection, is an important factor in the high metallicities and SFRs
we observe. As such, it is unlikely that the SFRs presented here are
applicable to all \CaII-selected galaxies. This is probably also true
of the super-solar metallicities measured, though the slope and
scatter in the luminosity--metallicity relationship ensures that the
effect should be less pronounced. Further observations of a larger,
less biased sample of \CaII-selected galaxies is required to confirm
this. A larger sample would also allow identification of general
trends in \CaII-selected galaxy properties, such as the relationships
between luminosity, SFR, metallicity, impact parameter, morphology
etc.

It may be possible to estimate this luminosity bias by comparison with
the results of \citetalias{Wild_etal_2007a} who derived an \OII-based
SFR from stacked SDSS spectra of higher-redshift \CaII\
absorbers. However, such a comparison is only possible if one assumes
that \citetalias{Wild_etal_2007a}'s star-formation signal originates
from some fraction of \CaII-selected galaxies with low impact
parameters and $\sim\!L^\ast$ luminosities. Another plausible scenario
(and one favoured by \citetalias{Wild_etal_2007a}) is that most strong
\CaII\ absorbers host low levels of \emph{in situ} star formation. In
this case, the luminosity bias in our sample cannot be determined but
it clearly must be large since we do observe some highly star forming
galaxies. Future observations of a more representative \CaII-selected
galaxy sample will focus on quantifying the relative importance of
these scenarios.

Reddening corrections and proper treatment of the stellar absorption
at $\Hb$ and $\Ha$ proved vital for a robust measurement of the SFRs
in our \CaII-selected galaxies. If no dust extinction corrections were
applied, our SFRs would be a factor of 20 smaller and would be
consistent with the low SFRs reported in previous low-$z$
emission-line studies of DLA host galaxies that did not apply a dust
correction. It is therefore likely that previous conclusions about the
apparently low SFRs in DLA-selected galaxies should be
revised. Indeed, it is probable that the majority of such galaxies
studied, via their emission-lines, to date are actively star-forming.

The observed slow increase in DLA metallicity from high redshifts to
the current epoch has raised concerns that DLAs may not be reliable
tracers of star formation in the Universe. However, the fact that we
find high emission-line metallicities and SFRs in \CaII-selected
galaxies at low-$z$ is consistent with previous studies which conclude
that DLAs can be closely linked to star formation without themselves
having high absorption-line metallicities. A study comparing
absorption- and emission-line metallicities as a function of impact
parameter would be an excellent way to assess this more directly; it
would constrain the relative contributions of dust and metallicity
gradients to any observed discrepancies. A handful of individual DLA
galaxies have already been studied in emission and absorption
\citep{Bowen_etal_2005a,Schulte-Ladbeck_etal_2004a,Chen_etal_2005a},
but a more complete sample of such galaxies is required for a
comprehensive study. Since \CaII\ absorbers and their host galaxies
are relatively easy to identify at $z<0.5$, they will be the ideal
targets for such a study when the community next has access to a
space-borne UV spectrograph. Those observations would also be vital
for quantifying the (likely strong) overlap between the \CaII\ and DLA
absorber populations.

\section*{Acknowledgments}

We are grateful to \'{A}ngeles D\'{i}az, Rob Kennicutt and John
Moustakas for their expert advice with the measurement of the spectra
and the interpretation of the galaxy morphologies. Dawn Erb and Jason
Melbourne kindly helped with the preparation of Figs.~\ref{fig:AGN} \&
\ref{fig:Z-L}. BJZ gratefully acknowledges the support of the UK
Particle Physics and Astronomy Research Council (PPARC). MTM thanks
PPARC for an Advanced Fellowship. We thank our anonymous referee for
helpful feedback. This work was based on observations collected at the
European Southern Observatory, Chile, as part of program 77.A-0756.

Funding for the SDSS and SDSS-II has been provided by the Alfred P.
Sloan Foundation, the Participating Institutions, the National Science
Foundation, the U.S. Department of Energy, the National Aeronautics
and Space Administration, the Japanese Monbukagakusho, the Max Planck
Society, and the Higher Education Funding Council for England. The
SDSS Web Site is \href{http://www.sdss.org/}{http://www.sdss.org/}.



\bspsmall

\label{lastpage}

\end{document}